\documentclass[final,5p,times,twocolumn]{elsarticle}

\usepackage{graphicx}
\usepackage{amssymb}
\usepackage{lineno,hyperref}
\graphicspath{{figures/}}

\journal{Nuclear Instruments and Methods in Physics Research Section A}

\usepackage{textcomp}
\usepackage{booktabs}
\usepackage{xfrac}
\usepackage{tikz}
\usetikzlibrary{shapes.geometric, arrows, automata, positioning}
\usepackage{comment} 
\usepackage{xcolor}
\tikzstyle{accelerator} = [rectangle, minimum width=0.5 cm, minimum height=2 cm, text centered, draw=black]
\tikzstyle{production} = [rectangle, rounded corners, minimum width=2cm, minimum height=2cm,text centered, draw=black]
\tikzstyle{valve} = [rectangle, minimum width=1cm, minimum height=1cm, text centered, text width=1cm, draw=black]
\tikzstyle{measurement} = [diamond, minimum width=1cm, minimum height=1cm, text centered, draw=black]
\tikzstyle{pump} = [circle, minimum size=1cm, text centered, draw=black]
\tikzstyle{arrow} = [thick,->,>=stealth]

\begin{document}

\begin{frontmatter}

\title{$^{23}$Ne Production at SARAF-I}


\author[SARAF,HUJI]
{
Yonatan Mishnayot
\corref{mycorrespondingauthor}
}
\cortext[mycorrespondingauthor]{Corresponding author}
\ead{yonatan.mishnayot@mail.huji.ac.il}

\author[HUJI]
{
Hitesh Rahangdale
}
\author[HUJI,ETH]
{
Ben Ohayon
}
\author[SARAF]
{
Sergey Vaintraub
}
\author[SARAF]
{
Tsviki Hirsh
}
\author[SARAF]
{
Leo Weismann
}
\author[SARAF]
{
Amichay Perry
}
\author[SARAF]
{
Asher Shor
}
\author[SARAF]
{
Arik Kreisel
}
\author[SARAF]
{
Shadi Ya'akobi
}
\author[SARAF]
{
Einat Buznach
}
\author[HUJI]
{
Guy Ron
}

\address[SARAF]{Soreq Nuclear Research Center, Yavne, 8180000}
\address[HUJI]{The Racah Institute of Physics, The Hebrew University of Jerusalem, Givat Ram, Jerusalem, 9190401}
\address[ETH]{Institute for Particle Physics and Astrophysics, ETH Z\"urich, CH-8093 Z\"urich, Switzerland}

\begin{abstract}
In this article, we present a measurement of flow rate, yield and effusion time of a $^{23}$Ne production and transport system. 
We used an accelerator-driven Li(d,n) neutron source to produce neutrons up to 20 MeV. 
The radioactive atoms were produced by a $^{23}$Na(n,p) reaction at a NaCl target. 
Later, the atoms were diffused out from the NaCl crystals and effused from the production chamber via a 10 m hose to a measurement cell and their decay products were detected using high purity germanium (HPGe) and plastic scintillator detectors. 
The resulting flow rate was $6.9\pm0.5\cdot 10^4\sfrac{atoms}{sec}$ and the total yield was $3.2\pm0.4\cdot10^{-9}\sfrac{atoms}{deuteron}$. We summarize our methods and estimates of efficiencies, rates of production and effusion.
\end{abstract}


\end{frontmatter}

\section{Introduction} \label{introduction}


Most of the reactors that currently produce radioisotopes are expected to come offline in the next few years \cite{2014-PRod}. This makes radioisotope production at accelerator facilities, which are relatively cost-effective and easy to operate, a growing field with applications in nuclear medicine \cite{1995-Med,2014-Zhuikov}, industry \cite{2012-Hamm} and basic science \cite{Abel19}.
In basic science, one such application is the search for Beyond Standard Model (BSM) physics in the weak interaction, through precision measurements of decay parameters such as the beta-neutrino angular correlation coefficient \cite{2018-Gonzalez}. This observable is predominantly measured today using ion or atom traps \cite{2019-Shidling, 2020-Omalley, ohayon2020}.


In 2018, we began operating the trapping laboratory at the Soreq Applied Research Accelerator Facility (SARAF) \cite{Ohayon18, Mardor18}. 
Two dedicated trap systems were commissioned to measure $\beta$ decay parameters: 
an Electrostatic Ion Beam Trap (EIBT) for $^6$He \cite{2011-EIBT, Mukul18}, and a Magneto-Optical Trap (MOT) for neon isotopes \cite{2015-ZS,2019-Imaging}. 
Since both systems involve short-lived isotopes (\textless 1 min), they have to be coupled to an accelerator-based radioisotope production and transport apparatus. 
Laser trapping of noble gas atoms is inefficient due to the need to excite them to a meta-stable state \cite{2003-Birkl,2015-RF}.
To obtain enough statistics from roughly $10^7$ detected events \cite{2016-OscarBeta,ohayon2020} within a reasonable beam time, one needs to produce these atoms in ample quantities \textgreater$10^8\sfrac{atoms}{sec}$.
Only high-current accelerators can produce enough atoms to feed the trap.


The following article will present the first application of the SARAF-I accelerator for mass production of radioisotopes using fast 
neutrons. 
We chose to focus on $^{23}$Ne, as only neon isotopes will be trapped in the MOT as part of the trapping program for testing the SM, and $^{23}$Ne is the longest-lived and the easiest to produce. 

The SARAF-I accelerator is designed to provide proton and deuteron beam currents of up to 2 mA. 
SARAF can also serve as an intense neutron source using its Liquid Lithium Target (LiLiT) \cite{2009-LiLIT, 2018-LILIT}. 
A deuteron beam of 5 MeV produces neutrons of up to 20 MeV via the $^7$Li(d,n) reaction. 
At a full deuteron beam, the production rate of $^{23}$Ne atoms is expected to be higher than $10^8\sfrac{atoms}{sec}$.

BeO and NaCl targets were designed and built for production of radioisotopes by neutron-induced nuclear reactions: $^9$Be(n,$\alpha$)$^6$He and $^{23}$Na(n,p)$^{23}$Ne, respectively.
The half-lives of the radioisotopes are 806.89 ms \cite{2012-Knecht} and 37.148 sec \cite{2015-Laffoley}, respectively.
The BeO target was made of porous lattice, while the NaCl target was made of powder, allowing better diffusion. 
In addition, both targets were heated to improve the diffusion from the crystal lattice.
As both helium and neon are noble gases, they may be effused through a vacuum hose.
We chose to design a simple vacuum transport system rather than using other methods that employ complex separation methods \cite{Abel19}. 

The current experiment follows our group's previous work at the Weizmann Institute \cite{Tom}. 
The current experiment main goal was to design an efficient $^{23}$Ne production and transport apparatus.
A prototype system, composed of a $^{23}$Na target, 10 m hose and a measurement cell, was constructed for this purpose.
The target was irradiated with fast neutrons to produce $^{23}$Ne atoms. 
The $^{23}$Ne atoms diffused outside the crystal lattice and effused to the measurement cell. 
An array of $\beta$ and $\gamma$ detectors was placed around the measurement cell and used to measure the decays, identify contamination and estimate the quantity of $^{23}$Ne atoms produced and effused. 
We describe the system in the second section, and discuss the experimental results in the third and fourth sections.


\section{Methods} \label{methods}
The three main methods for $^{23}$Ne production are:
\begin{itemize}
    \item Deuteron-induced reactions such as $^{22}$Ne(d,p) at 2.6 MeV \cite{1940-Watson} and $^{23}$Na(d,2p) at 22 MeV \cite{Penning57}.
    \item Thermal neutron-induced reaction $^{22}$Ne(n,$\gamma$)
    on enriched neon gas \cite{Moreh68, Lancman65}. 
    \item Fast neutron-induced reaction $^{23}$Na(n,p) on different sodium compounds \cite{1937-Bjerge, 1958-Allen, 1958-Ridely}.
\end{itemize}
Both deuteron-induced reactions and thermal neutron-induced reactions were eliminated for the following reasons: 
A $^{22}$Ne gas target involve effusion of stable neon gas throughout the system, effectively increasing the gas load in trap and reducing trap lifetime;
Deuteron-induced reactions complicate the target design, which usually required a cooled target window. 
In addition, the $^{23}$Na(d,2p) reaction at 22 MeV is above SARAF-I beam capabilities.
In order to avoid effusion of stable neon gas in the system, we chose the fast neutron reaction $^{23}$Na(n,p) with threshold of 4 MeV \cite{1961-Williamson}.
Since SARAF maximal deuteron energy is 5.6 MeV, resulting in neutrons up to 20 MeV at neutron source \cite{Tsviki, 2015-SARAF}, the choice of $^{23}$Na(n,p) reaction is optimal for SARAF.
In addition, using $^{23}$Na(n,p) reaction simplifies the target design, avoiding the need for a special target window.
NaCl was chosen as a target material due to its stability, ease of use and safety.

The LiF Thick Target (LiFTiT) \cite{2012-LiFTiT} was connected to the end of the SARAF beamline. 
The LiFTiT was made of 300 $\mu$m thick LiF crystals glued to a water-cooled copper flange. 
The surface of the crystals was painted with a thin layer of carbon paint to avoid build up of the beam charge. 
The flange was connected to the final section of the beam line via a 20 cm pipe and Teflon sealing, allowing direct collection of beam charge from the electrically insulated flange. 
The long connected pipe allowed efficient collection of secondary electrons. 
A pair of permanent magnets were set near the target flange for secondary suppression of electrons, and the accelerator tune was performed with the Rutherford backscattering monitor \cite{2009-Weissman}.
The final diagnostic station used for beam tuning was placed half a meter upstream from the LiF target. 
The beam duty cycle was 1.4\% and peak beam value was 250 $\mu$A, corresponding to an average beam current of 3.4 $\mu$A. 
A neutron dosimeter monitor was placed 3 m from the LiFTiT to monitor neutron dose rate and the stability of the LiFTiT. 
As the LiF crystals could not sustain significantly higher current, they were eventually burnt out towards the end of the experiment.

\begin{figure}[ht]
    \centering
    \includegraphics[width = 0.45\textwidth, trim={0.5cm 1cm 5.5cm 1cm},clip]{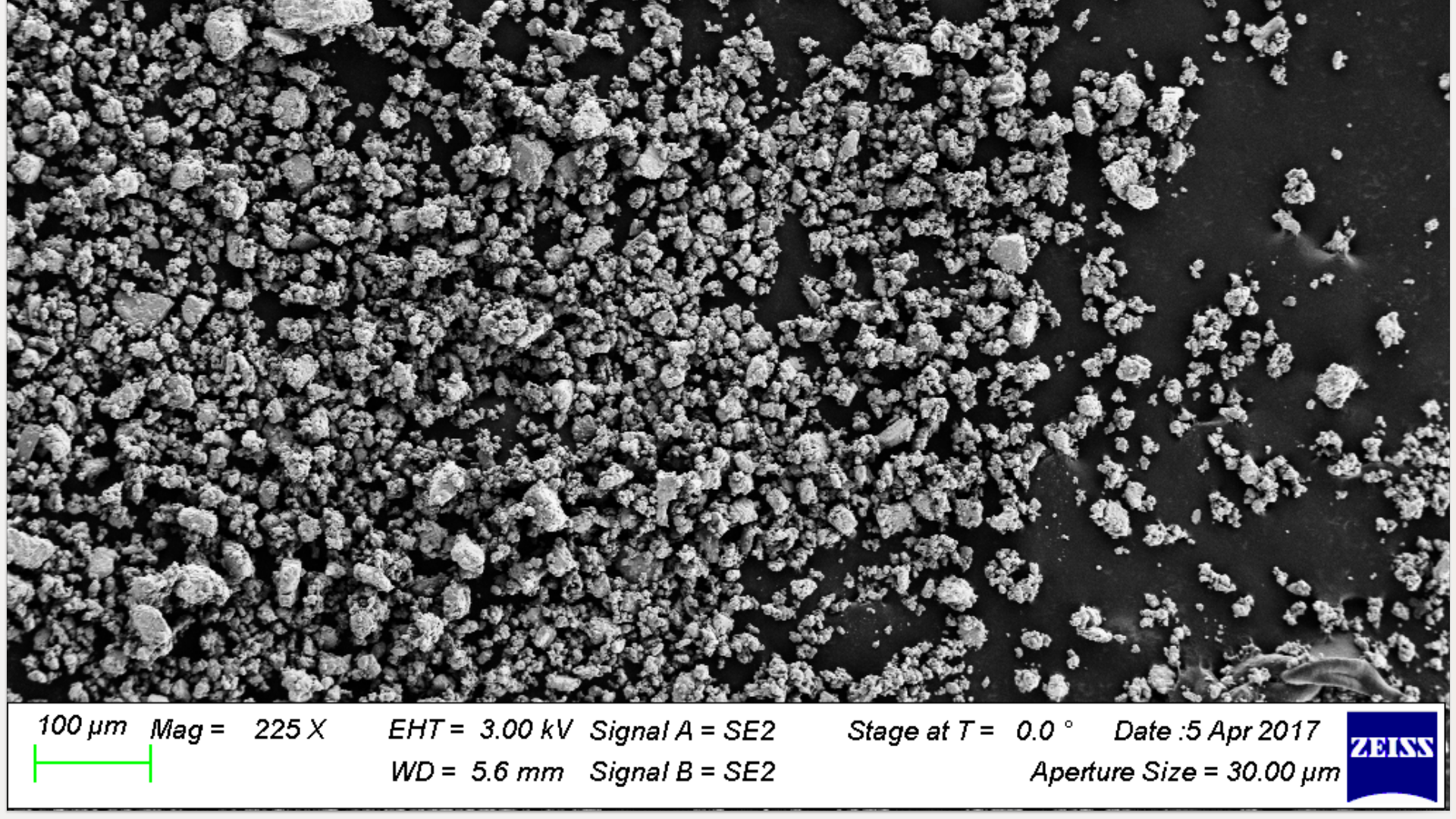}
    \caption{SEM scan of a salt sample, used to estimate the crystal's size.}
    \label{fig:grain size}
\end{figure}

The target consisted of 1.4 kg of a ground table salt (NaCl) that was sifted to a crystal size smaller than 40 $\mu$m. 
To verify the crystal size, several samples of the sifted salt were measured using scanning electron microscopy (SEM) (fig. \ref{fig:grain size}).
The SEM scans were analyzed using MIPAR software \cite{2017_Sosa}, which showed that the crystal size is exponentially distributed, with the average being $10\ \mu$m.
The salt was stored in a production chamber made of stainless steel half nipple (4" diameter, 32.5 cm height), wrapped in heating tape, glass wool and high-temperature aluminum tape. 
The production chamber was placed in front of the LiFTiT, and connected to a hose 10 m in length and 3" in diameter, via a 4" gate valve that separated the production chamber from the hose.
This separation enabled us to measure the effusion time in the experiment.
Prior to experiment, the production chamber was heated up to 380 \textdegree C to remove water vapors.
During the experiment, the production chamber was heated to an average temperature of $360\pm 20$\textdegree C to improve the diffusion of $^{23}$Ne atoms out of the salt crystals.

\begin{figure}[ht]
    \centering
    \includegraphics[width=.9\linewidth]{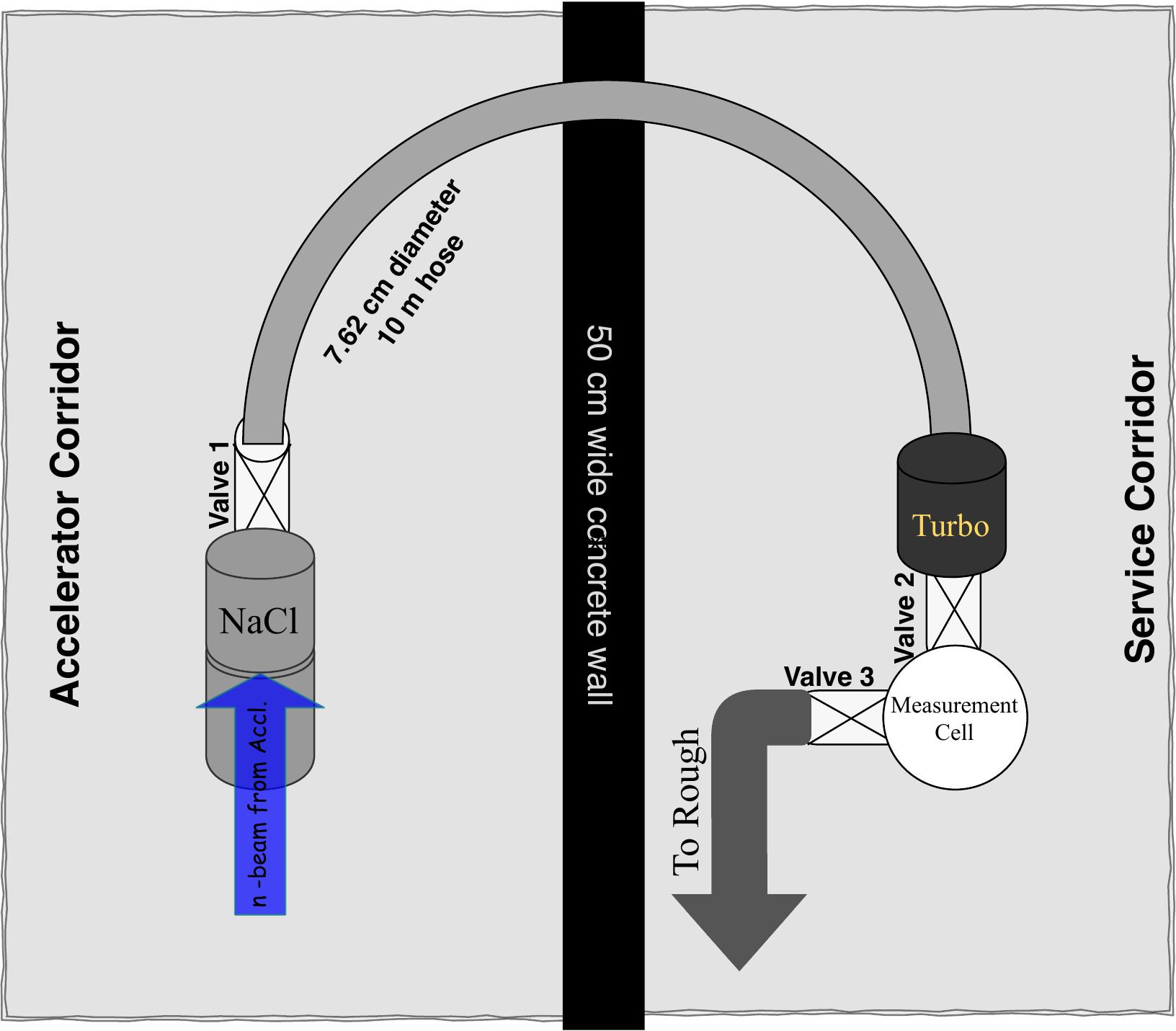}
    \caption{\small{A sketch of the experimental system. The measurement cell setup is shown in detail in fig. \ref{fig:System}.}}
    \label{fig:Scheme}
\end{figure}

The hose end was coupled to a $520\sfrac{l}{s}$ turbo-molecular pump, whose outlet was connected to a 6 cm diameter (3.8 cm height) measurement cell via a 1" valve. 
Another 1" valve was connected the other side of the measurement cell to a roughing pump, which was pumped out to an exhaust line.
Both valves -- in the measurement cell inlet and outlet(valves 2 and 3 in fig. \ref{fig:Scheme})  -- were used to control the flow into the measurement cell and the accumulation of $^{23}$Ne atoms in the cell. Fig. \ref{fig:Scheme} shows a schematic view of the experimental system.

\begin{figure}[ht]
    \centering
    \includegraphics[width=0.9\linewidth]{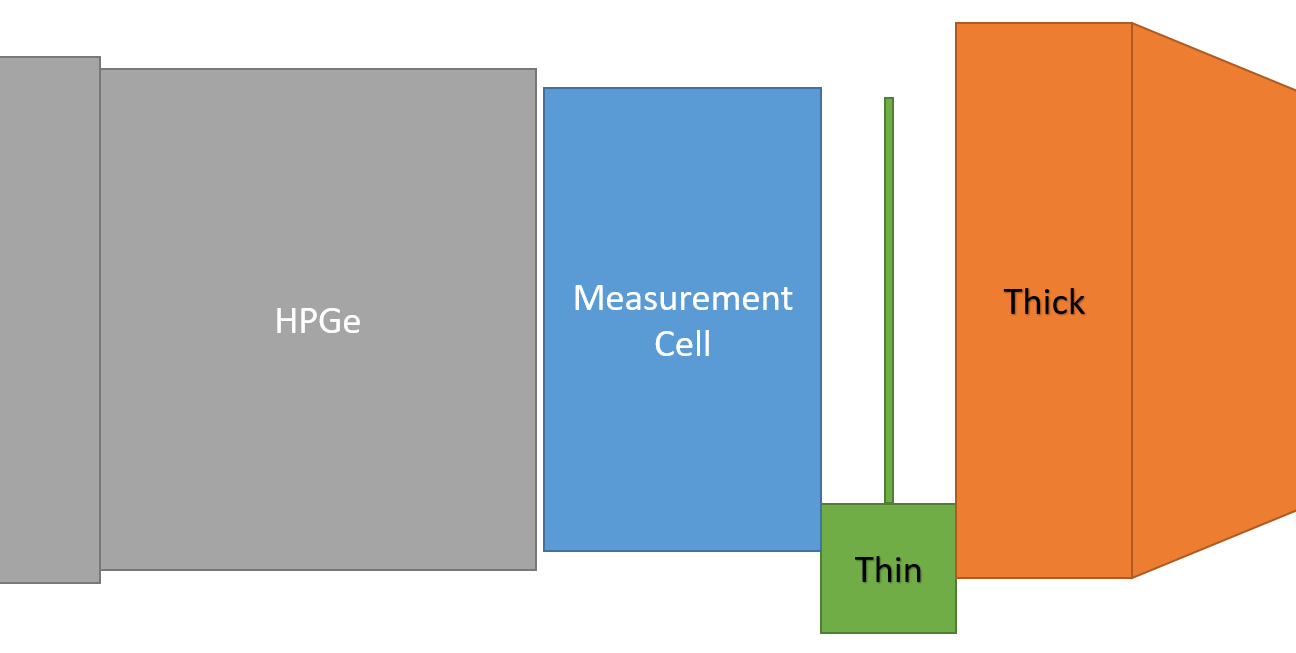}
    \caption{\small{A sketch of the measurement system and the detectors: Leftmost---HPGe, with the  measurement cell to its right.
    Rightmost---the thick scintillator, with the thin detector to its left.}}
	\label{fig:System}
\end{figure}

The measurement cell iteslf had a 2 mm stainless steel window in front and a 25 $\mu$m Mylar window in back.
A p-type HPGe (Ortec GEM40-83) detector was placed in front of the stainless steel window to measure $\gamma$ rays.
Two plastic scintillators were placed in front of the Mylar window as a $\Delta E - E$ telescope. 
The telescope was made of a thin, square scintillator (0.5 mm thickness, 5 cm side length) and a thick, cylindrical scintillator (2.5 cm thickness, 8.8 cm diameter).
Any particle that left a signal in both scintillators was tagged as $\beta$ particle and its energy measured by the thick scintillator. 
Fig. \ref{fig:System} shows schematically the measurement cell and detectors.

\section{Flow and Yield} \label{steady state measurements}

In this section, we discuss the experimental results obtained in long measurements ($\sim$ 30 min).
During the measurements we left both valves 1 and 2 (presented in fig. \ref{fig:Scheme}) open, while valve 3 was closed and opened to empty the measurement cell when its pressure increased to 2-3 Torr (1--3 times during a single measurement). 
In the first subsection, we discuss the yield and flow-rate evaluation of the long measurements data.
In the second and third subsections, we discuss two additional results found in the long measurement data:
the diffusion of $^{23}$Ne atoms out of the salt crystals and the detection of other radioisotopes in the measurement cell.


\subsection{Flow Rate and Yield} \label{flux and yield}

\begin{figure}[ht]
    \centering
	\includegraphics[width=\linewidth]{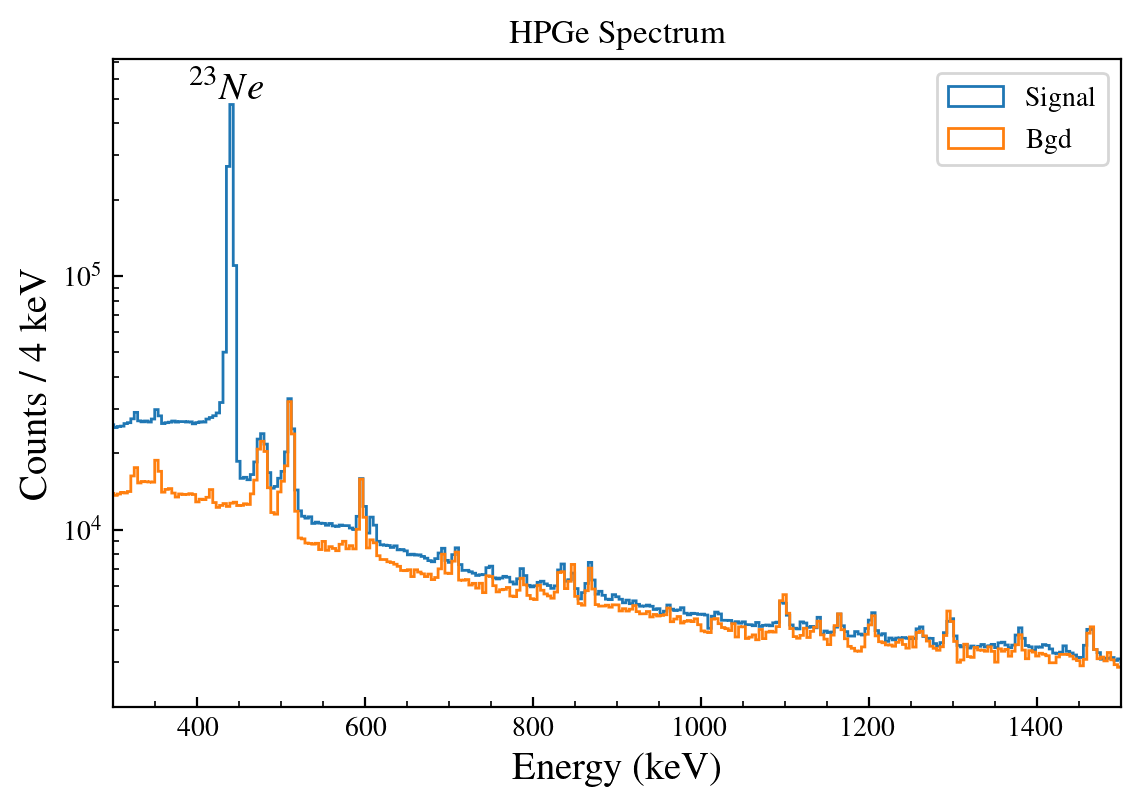}
	\caption{\small{Zoom in on a HPGe spectrum, focusing the 440 keV $^{23}$Ne peak. The spectrum was compared to the normalized accelerator's background. The background was taken when the accelerator was on, but valve no. 1 in fig. \ref{fig:Scheme} was closed}}
		\label{fig:Ne23Spectrum}
\end{figure}

\begin{figure}[ht]
    \begin{center}
        \resizebox{8 cm}{8 cm}{
            \begin{tikzpicture}[auto]
            \draw [line width = 0.05cm] (0,7) -- (1.9,7);
            \draw [line width = 0.05cm] (0,7) -- (0,1);
            \draw [arrow] (0,4.2) -- (1.9,4.2);
            \draw [arrow] (0,3) -- (1.9,3);
            \draw [arrow] (0,1.7) -- (1.9,1.7);
            \draw [arrow] (0,1) -- (1.9,1);
            \draw [line width = 0.05cm] (2.0,4.2) -- (7.1,4.2);
            \draw [line width = 0.05cm] (2.0,3) -- (7.1,3);
            \draw [line width = 0.05cm] (2.0,1.7) -- (7.1,1.7);
            \draw [line width = 0.05cm] (2.0,1) -- (7.1,1);
            \draw [arrow] (3.0,4.2) -- (3.0,1);
            \draw [arrow] (3.4,4.2) -- (3.4,1.7);
            \draw [arrow] (3.6,3) -- (3.6,1);
            \draw [arrow] (4.0,3) -- (4.0,1.7);
            \draw [arrow] (4.2,1.7) -- (4.2,1);
            \node [text width=1cm] at (0.5, 7.2) {\small{$5/2+$}};
            \node [text width=1cm] at (2.5, 4.4) {\small{$3/2+$}};
            \node [text width=1cm] at (2.5, 3.2) {\small{$7/2+$}};
            \node [text width=1cm] at (2.5, 1.9) {\small{$5/2+$}};
            \node [text width=1cm] at (2.5, 1.2) {\small{$3/2+$}};
            \node [text width=3cm] at (1.6, 6.0) {$Q_{\beta-}=4375.81(10)$};
            \node [text width=2cm] at (1.4, 7.6) {$37.148$ $s$};
            \node [text width=1cm] at (6.4, 4.4) {\small{$2982.060$}};
            \node [text width=1cm] at (6.4, 3.2) {\small{$2076.011$}};
            \node [text width=1cm] at (6.5, 1.9) {\small{$439.990$}};
            \node [text width=1cm] at (7.4, 1.2) {\small{$0$}};
            \node [text width=1cm] at (7.65, 4.2) {\small{$2.5$ $fs$}};
            \node [text width=1cm] at (7.65, 3.0) {\small{$24$ $fs$}};
            \node [text width=1cm] at (7.65, 1.7) {\small{$1.24$ $ps$}};
            \node [text width=1cm] at (7.65, 1.05) {\small{$stable$}};
            \node [text width=5cm, rotate = 60, fill=white] at (4.2, 6.55) {\small{$58.79$ $2981.85$ $M1(+E2)$}};
            \node [text width=5cm, rotate = 60, fill=white] at (4.6, 6.55) {\small{$41.21$ $2541.92$ $M1+E2$}};
            \node [text width=5cm, rotate = 60, fill=white] at (4.8, 5.35) {\small{$8.2$ $2075.91$ $E2(+M3)$}};
            \node [text width=5cm, rotate = 60, fill=white] at (5.2, 5.35) {\small{$91.8$ $1635.96$ $M1+E2$}};
            \node [text width=5cm, rotate = 60, fill=white] at (5.4, 4.05) {\small{$100$ $439.986$ $M1+E2$}};
            \node [text width=1cm] at (0.6, 4.4) {\small{$0.065\%$}};
            \node [text width=1cm] at (0.6, 3.2) {\small{$1.1(1)\%$}};
            \node [text width=1cm] at (0.6, 1.9) {\small{$32(3)\%$}};
            \node [text width=1cm] at (0.6, 1.2) {\small{$67(3)\%$}};
            \node [text width=1cm] at (2.2, 7.2) {\small{$0$}};
            \node [text width=1cm] at (1.8, 4.4) {\small{$6.13$}};
            \node [text width=1cm] at (1.8, 3.2) {\small{$5.82$}};
            \node [text width=1cm] at (1.8, 1.9) {\small{$5.38$}};
            \node [text width=1cm] at (1.8, 1.2) {\small{$5.27$}};
            \node [text width=2cm] at (1.6, 6.5) {\textcolor{green}{\large{$^{23}_{10}$Ne}}};
            \node [text width=2cm] at (5.0, 0.5) {\textcolor{green}{\large{$^{23}_{11}$Na}}};
            \end{tikzpicture}
        }
    \caption{The decay scheme of $^{23}$Ne.}
    \label{fig:23BR}
    \end{center}
\end{figure}
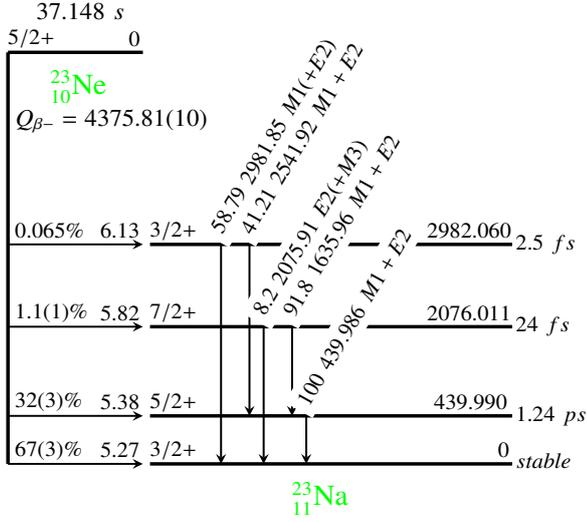

In order to calculate the flow rate into the measurement cell, we used the major $\gamma$ peak of $^{23}$Ne at 440 keV obtained by HPGe (fig. \ref{fig:Ne23Spectrum}). 
We took a histogram of the events' time stamps  and fitted it to the expected flow, described by:
\begin{equation}
    \lambda N = K + C \cdot exp(-\lambda t) 
\end{equation}
where $K$ is flow rate at longer times -- after several minutes, $\lambda N$ is number of decays in the measurement cell, $C$ is constant and $\lambda = ln2/t_{1/2}$ is the decay time constant.
We considered several factors during the evaluation of decays in the measurement cell:
the HPGe efficiency;
the branching ratio (BR) of $^{23}$Ne (see fig. \ref{fig:23BR});
and losses of $\gamma$ rays in the stainless steel window. 
In addition, we had to correct the volume of the measurement cell, as the atoms were spread throughout the volume between the pump outlet and the last valve, while we measured only the decays in front of the window. 
We found the resulting flow rate to be  $6.9\pm0.5\cdot 10^4 \sfrac{atoms}{sec}$. 
Once we had calculated the flow rate and the beam current, we evaluated the yield easily by dividing the flow rate by the number of deuterons in the beam. 
We found that the total yield was $3.2\pm0.4\cdot10^{-9}\sfrac{atoms}{deuteron}$.

\subsection{Diffusion} \label{diffusion}

In order to evaluate the diffusion out of the salt crystals, we had to calculate the effusion efficiency throughout the system, making it easy to exclude the effusion and focus only on the diffusion process.
To calculate the effusion efficiency, we simulated it in Molflow, following the comparison in the next section \ref{dynamic measurments}.
We defined the effusion efficiency as the number of atoms that passed the turbo pump, divided by the number atoms emitted from the target.
In addition, we set the target to emit atoms continuously.  
Thereafter, we used Molflow to simulate the effusion efficiency of radioactive $^{23}$Ne with a half life time of 37.15 sec and found the effusion efficiency to be 91\%.
Considering the $^{23}$Ne flow rate into the  measurement cell (as calculated in section \ref{flux and yield}) and the effusion efficiency, we calculated that the $^{23}$Ne flow rate at the production chamber outlet was $7.6\pm0.5\cdot 10^4\sfrac{atoms}{sec}$.

After obtaining the $^{23}$Ne flow rate in the production chamber outlet, we had to find the $^{23}$Ne production rate in the salt target.
Due to the high neutron flux on the salt target, we could not directly measure the $^{23}$Ne production rate in the target. 
Therefore, we simulated it using GEANT4 \cite{Geant4a, Geant4b, Geant4c} as follows:
First, we inserted the neutron emission spectrum of 5 MeV deuteron beam on LiFTiT \cite{Tsviki, 2015-SARAF} into GEANT4. 
Then we simulated the $^{23}$Na(n,p)$^{23}$Ne reaction in the salt target using GEANT4.
Normalizing the results to the beam current, we found that the production rate was 
$1.7\pm0.3\cdot 10^6 \sfrac{atoms}{sec}$.
Having both $^{23}$Ne production rate and flow rate at the production chamber outlet, we divided the flow rate by the production rate to get a diffusion efficiency of $4.5\pm0.5\cdot 10^{-2}$, equivalent to a diffusion time of $166\pm8$ sec.
Using the diffusion time and the averaged temperature, we calculated the average crystal size as a sanity check.
Our calculation is based on the calculation in \cite{Tom}. 
Assuming cubic crystals, we calculated the average crystal size to be $47\pm15\ \mu$m.
Although that result is not consistent with the average crystal size obtained by SEM (in the Methods section \ref{methods}), it is very close to the upper limit obtained by the sieve of $40\ \mu$m.
Therefore, given the uncertainties of temperature and diffusion time, the sanity check justifies our results.

\subsection{Contaminations} \label{contamination}

Throughout the experiment, we were surprised to find radioisotopes other than $^{23}$Ne in the measurement cell. 
Generally, due to the inertness of noble elements, we would expect their atoms to diffuse as mono-atomic molecules out of the salt crystals and to effuse throughout the system to the measurement cell. 
In contrast, we expected atoms of other elements to be trapped in the crystal lattice or be absorbed by the system walls. 

\begin{figure}[ht]
    \centering
	\includegraphics[width=\linewidth]{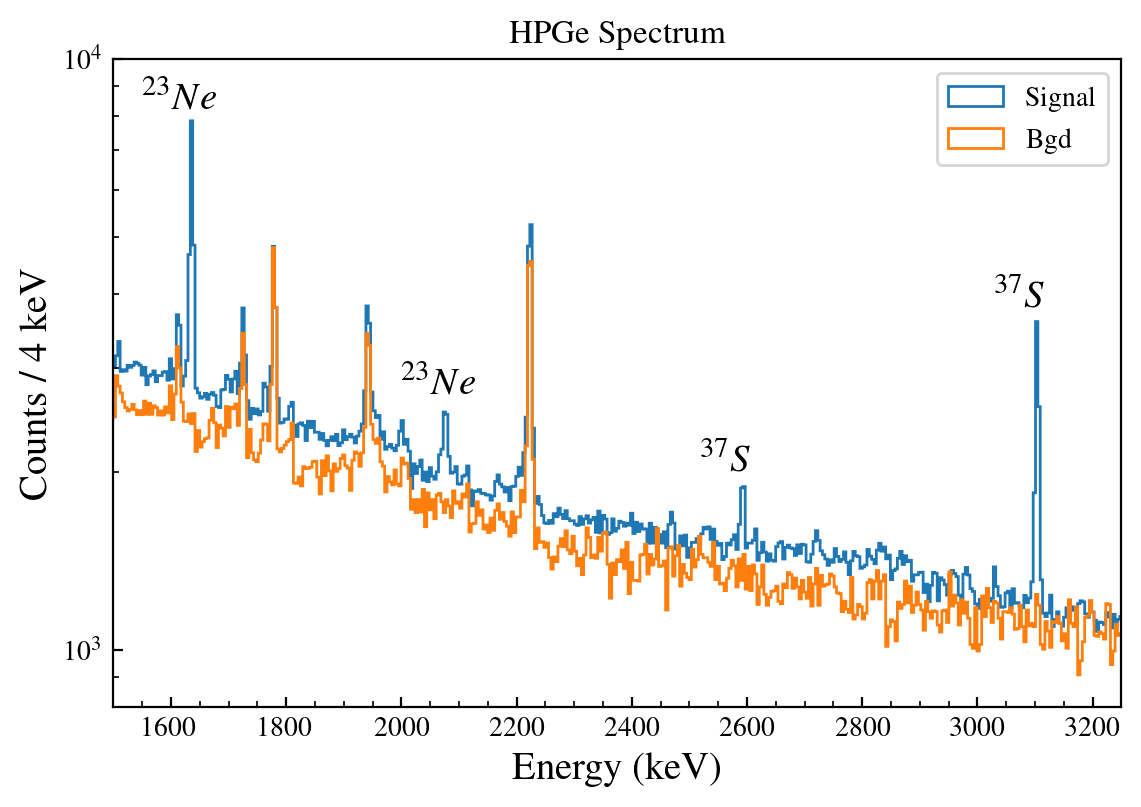}
	\caption{\small{Zoom in on a HPGe spectrum, focusing on the 1636 and 2076 keV $^{23}$Ne peaks and the $^{37}$S peaks. The spectrum is compared to a normalized accelerator background. The unlabeled peaks are activation peaks of the HPGe. The background was taken when the accelerator was on, but valve no. 1 in fig. \ref{fig:Scheme} was closed}}
	\label{fig:S37activation}
\end{figure}

During the experiment we detected clear traces of other elements, particularly $^{37}$S.
Since the contribution of $^{37}$S was negligible ($\sim 1\%$), we ignored it and considered only $^{23}$Ne effusion.
In the HPGe spectrum, we found the 3103 keV $^{37}$S $\gamma$ peak and its escape peak at 2592 keV (in fig. \ref{fig:S37activation}). 
The $^{37}$S is produced in the $^{37}$Cl(n,p)$^{37}$S reaction, with a threshold of 11 MeV \cite{1979-vanska}, and its half life time is 5.05 min.
We assume that due to its high vapor pressure at high temperatures, 10 kPa at 591 K \cite{West29}, the sulfur atoms or molecules diffused out of the salt crystals and reached the measurement cell.

Another possible contamination is $^{20}$F, produced by the $^{23}$Na(n,$\alpha$)$^{20}$F reaction with threshold of 6 MeV \cite{1983-Weigmann}.
Based on \cite{1956-Gerber}, we are almost certain that $^{20}$F was produced in the salt target. However, we wanted to confirm or reject the existence of $^{20}$F in the measurement cell.
Since the strongest $\gamma$ peak energy of $^{20}$F is 1633 keV, it interferes with the 1636 keV $^{23}$Ne peak.
On the other hand, we can use the 1636 keV peak to test for $^{20}$F in the measurement cell.
We calculated the resulting ratio of $^{23}$Ne 1636 and 440 keV peaks (the $^{23}$Ne peaks in figs. \ref{fig:Ne23Spectrum} and \ref{fig:S37activation}) as $2.98\pm0.11\%$.
Unlike targets made of $^{23}$Na, no $^{20}$F was produced in targets made of $^{22}$Ne. 
Therefore, we compared our calculated ratio to the previous results of $^{23}$Ne branching ratio measurements conducted using $^{22}$Ne target.
The previous results were $2.81\pm 0.06 \%$ \cite{Moreh68} and $3.03\pm0.12\%$ \cite{1974-Alburger}, while our result was $2.98\pm0.11\%$. 
Since our result is located between the previous results, we can reject the existence of $^{20}$F in the measurement cell. 
We assume that due to its high reactivity, $^{20}$F atoms were bound to $^{23}$Na atoms in the crystal lattice. 

\section{System Design} \label{dynamic measurments}

In the future, we plan to produce and convey several radioisotopes to our trapping systems, especially noble gases.
Since noble gases are inert, we would like to transport the atoms via effusion through vacuum pipes, and simulate the effusion through the vacuum transport system as part of its design process.
We therefore chose Molflow \cite{2019-Kersevan}, a dedicated simulation tool for vacuum systems, to simulate our system.
In order to validate our simulation results, we used short measurements with an output comparable to our simulation results. 
In the first subsection, we discuss the measurements independent of the simulations, while in the second subsection we describe the simulations and discuss their comparison to the measurements.

\subsection{Effusion} \label{effusion}
Since the number of $^{23}$Ne atoms is negligible compared to other gases in the system, we chose to measure its decays in the measurement cell. 
The measurement included several cycles, each lasting 150 sec, in which all cycles had the same effusion profile to maintain consistency. 
We started the cycles when the gate valve on top of the production chamber (valve 1 in system scheme, fig. \ref{fig:Scheme}) was closed  for 30 seconds. 
After 30 seconds, we opened the gate valve on top of the production chamber for 5 seconds and then closed it again.
After 150 seconds, we opened the valve to the rough pump and exhaust line (valve 3 in system scheme, fig. \ref{fig:Scheme}) to empty the cell. Valve 2 in fig. \ref{fig:Scheme} was left open during the whole measurement.

We used the $\Delta E - E$ telescope to tag the emitted $\beta$ particles throughout the cycle, focusing on time profile rather than energy and absolute yield.
The measurement results are shown in the blue dotted curve shown in the bottom right subfigure of the simulation and experiment comparison in fig. \ref{fig:flowcomp}.

\subsection{Molflow Simulations} \label{molflow simulations}

As described at the beginning of this section, we used Molflow to simulate the effusion throughout the system. 
More precisely, we used Molflow to simulate the effusion of $^{23}$Ne atoms from the production chamber outlet---where the atoms were created in simulation---to the measurement cell inlet, where the atoms were destroyed in simulation and where we measured the decays. 

While using Molflow, our main challenge was the effusion of particles through vacuum pumps.
In Molflow, vacuum pumps are sinks; i.e., any particle that goes in is destroyed. 
However, we were interested in particles that passed through the vacuum pumps. 
Rather than using pump elements in the simulation, we modeled the pump by inserting two planes into the pump body. 
The first plane was partially opaque toward the particle source at the production chamber outlet, while the second, 1 cm downstream, was fully opaque toward the particle sink at the measurement cell inlet.    
We calculated the first plane opacity to be 90\%. 

\begin{figure}[ht]
    \centering
    \includegraphics[width=0.95\linewidth]{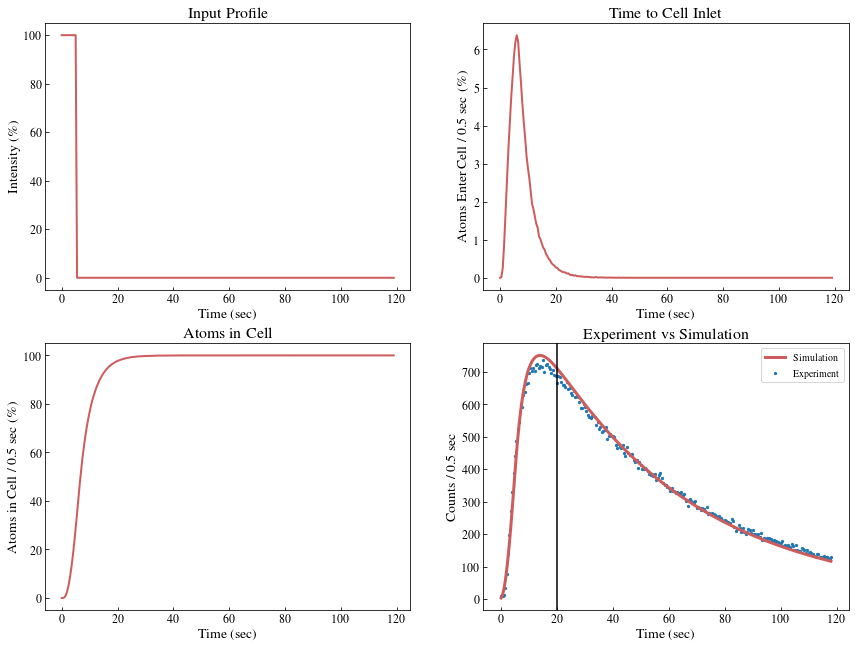}
    \caption{\small{The simulation process and comparison to experimental data (left to right). The first three figures (top row and bottom left) show the simulation process as described in the Molflow simulations subsection \ref{molflow simulations}, and the bottom right figure shows the comparison to experimental data after applying decay and normalizing, with the different regions separated by a vertical black line.}}
    \label{fig:flowcomp}
\end{figure}

We changed both pipe length and plane opacity values in steps of 25 cm and 2.5\% respectively (except the highest opacity value, where we used 99\% instead of 100\%, which would mean no effusion), to validate the assignment of the measured pipe length---10 m with a calculated plane opacity of 90\% in the simulation.
In addition to the measured pipe length and calculated plane opacity,
we simulated 4 points below and 4 points above for both parameters to get 81 simulations. 
Then we modified the simulation results and compared them to the experimental data.
The simulation process and its modification to compare it to experimental data is described below:
\begin{enumerate}
    \item We started with a surface source of $^{23}$Ne atoms on top of the salt target, and the source emitted atoms for 5 sec (fig. \ref{fig:flowcomp}, upper left). This source is equivalent to opening the gate valve on top of the production chamber for 5 sec.
    \item We simulated the effusion of $^{23}$Ne atoms to the measurement cell inlet as a function of time using Molflow (fig. \ref{fig:flowcomp}, bottom left).
    \item We accumulated the effused atoms to find the number of atoms in the  measurement cell as a function of time (fig. \ref{fig:flowcomp}, top right).
    \item We applied a radioactive decay with $t_{1/2} = 37.15$ sec to the modified simulation results. Then we normalized the simulation results with radioactive decay to the experimental data. (fig. \ref{fig:flowcomp}, bottom right).
\end{enumerate}

\begin{figure}[ht]
    \centering
	\includegraphics[width=0.95\linewidth]{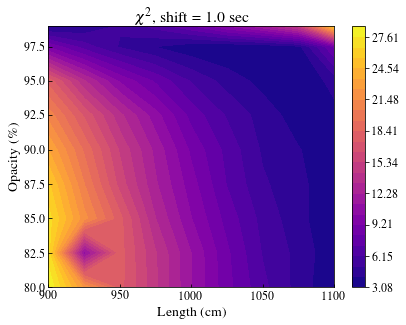}
	\caption{\small{A $\chi^2$ map at a shift of 1 sec, the minimal $\chi^2$ obtained for a length of 1025 and opacity of 97.5\%.}}
	\label{fig:ChiSquare}
\end{figure}

Since the opening time of the gate valve on top of the production chamber was not fixed, we did not have a well-defined experimental starting point to compare to our Molflow simulations.
Instead, we shifted the simulation results relative to the determined opening time in steps of 0.5 seconds up to 2.5 seconds. 
Then we compared the shifted simulation results to the experimental data. 
As clearly seen in the bottom right sub-figure of fig \ref{fig:flowcomp}, following the vertical black line, we can divide the time scale into two separate regions dominated by different processes. 
The effusion dominates the first, up to $\sim$ 20 seconds, while the decay dominates the second, from 20 seconds to the end of measurement.
Since we focused on the effusion time rather than the decay, we used only the first 20 seconds to calculate the $\chi^2$ value of the simulations at each shift (6 sets of 81 values, 486 $\chi^2$ values in total). 
We found the minimal $\chi^2$ value of 2.86 at shift of 1 sec, pipe length of 1025 cm and plane opacity of 97.5\%.
We can see that the deviation of the 'correct' simulated pipe length from the measured value is small and acceptable.
However, the deviation of the 'correct' simulated plane opacity from the calculated value is larger. 
Since the calculation of the plain opacity involves several parameters, we assume that deviation in some of the inputs increased its deviation.
The $\chi^2$ map at shift of 1 is shown in fig. \ref{fig:ChiSquare}.


\subsection{Future Systems} \label{future systems}
As mentioned previously, we plan to to produce and transport other radioisotopes to our trapping system. 
The next radioisotope to be trapped in our lab is  $^{6}$He. In contrast to  $^{23}$Ne, the $^{6}$He half life  is much shorter -- 0.807 ms.
Thus, we would like to estimate the yield of our transport system for a future use with $^{6}$He.
Following the comparison between our simulations to experimental results, we can reliably estimate the transport times and yield of our system for $^{6}$He atoms.
Table \ref{tab:percent} shows the estimated time took for a given fraction of the atoms to enter the measurement cell. 
the table includes both $^{23}$Ne and $^{6}$He, the estimated uncertainties are 0.5 and 0.1 seconds respectively.
\begin{table}[ht]
    \centering
    \resizebox{\linewidth}{!}{%
    \begin{tabular}{|c|c|c|c|c|c|c|c|}
    \hline
        Isotope & 1\% & 5\% & 10\% & 25\% & 50\% & 75\% & 90\% \\ \hline \hline
         $^{23}$Ne & 1.5 & 2.5 & 3.0 & 4.5 & 6.5 & 9.5 & 13.5 \\ \hline
         $^6$He & 1.0 & 1.6 & 2.2 & 3.4 & 5.0 & 6.6 & 8.6 \\ \hline 
    \end{tabular}
    }
    \caption{Comparison of transport times in seconds for $^{23}$Ne and $^6$He at different percentages, the uncertainties are 0.5 and 0.1 seconds respectively.}
    \label{tab:percent}
\end{table}

Another result that can be extracted using the simulation is the yield of our transport system for $^{6}$He.
Contrary to the time dependent measurements and simulations discussed in this section, we simulated the yield of a target emitting atoms continuously (in the Diffusion subsection \ref{diffusion}).
We divided the number of atoms that passed the turbo pump by the number of atoms emitted from the target.  
We found the effusion efficiency of  $^{6}$He to be 24\% compared to 91\% of $^{23}$Ne.

\section{Conclusions} \label{conclusions}

In this article, we presented the methods we used to calculate the flow rate, yield and effusion time of $^{23}$Ne in our prototype production and transport system. 
We validated our use of Molflow to simulate the effusion in vacuum systems by comparing it to experimental data of dynamic measurements.
Thereafter, we fitted of $^{23}$Ne $\gamma$ peaks at 440 and 1636 keV and used Molflow and GEANT4 simulations to extract the flow rate and yield.
Following the fittings and simulations, we found the resulting flow rate to be $6.9\pm0.5\cdot 10^4\sfrac{atoms}{sec}$ and the total yield to be $3.2\pm0.4\cdot10^{-9}\sfrac{atoms}{deuteron}$.
Additionally, we calculated the diffusion time out of salt crystals and found it to be $167\pm 8$ sec.

In this work, we showed that the diffusion dominates the system yield while the effusion effects are secondary.
Therefore, changing the transport line length by a meter or two will not change the flow rate significantly.
Wיhereas, reducing the diffusion time will improve the yield dramatically.
A comparison performed between $^{6}$H and $^{23}$Ne using Molflow simulations, also showed that the effusion contribution to the system yield is secondary, even in the case of short-lived isotopes. 

To meet the requirements of the MOT, we need to redesign the system to increase its flux production.
Following the conclusions, we need to consider to improve the diffusion by increasing the mass and the temperature \cite{Mehrer07, Askeland11} of the salt in the target.

We have already designed a new production chamber following these factors. 
The new production chamber includes three arms, each of which contains 1.6 kg of salt and can be heated up to 600$^{\circ}$C. 
In preliminary measurements of the new production chamber, taken during the last year, scaled following the methods described in this article to SARAF-II, and given higher beam energies with current of 1 mA at 100\% duty cycle - producing neutrons at averaged energy of 10-15 MeV - we estimate the $^{23}$Ne flux to be \textgreater$ 10^8\sfrac{atoms}{sec}$.
Therefore, we expect the new production chamber to fulfill the MOT requirements, feed the MOT with sufficient flux and allow experiments to be completed in a reasonable time.

\section{Acknowledgement}

The work presented here is supported by grants from the Pazy Foundation (Israel), and the Israel Science Foundation (grants no. 139/15 and 1446/16). This project has also received funding from the European Research Council (ERC) under the European Union’s Horizon 2020 research and innovation programme (Grant agreement No.  714118 TRAPLAB). BO is supported by the Ministry of Science and Technology, under the Eshkol Fellowship.
We thank the technical team of SARAF and the Weizmann Institute, especially Danny Kijel, Yigal Schahar and Gedalia Perelman, who planned the measurement cell and Dr Ofer Aviv who helped us with calibration sources. This work was started by Prof. Micha Hass, who passed away before it could be completed. 
This paper is dedicated to his memory.











\bibliographystyle{model1-num-names.bst}
	\bibliography{Ne23Bib.bib}

\begin{thebibliography}{47}
\expandafter\ifx\csname natexlab\endcsname\relax\def\natexlab#1{#1}\fi
\providecommand{\bibinfo}[2]{#2}
\ifx\xfnm\relax \def\xfnm[#1]{\unskip,\space#1}\fi
\bibitem[{Starovoitova et~al.(2014)Starovoitova, Tchelidze, and
  Wells}]{2014-PRod}
\bibinfo{author}{V.~N. Starovoitova}, \bibinfo{author}{L.~Tchelidze},
  \bibinfo{author}{D.~P. Wells},
\newblock \bibinfo{title}{Production of medical radioisotopes with linear
  accelerators},
\newblock \bibinfo{journal}{Applied Radiation and Isotopes}
  \bibinfo{volume}{85} (\bibinfo{year}{2014}) \bibinfo{pages}{39 -- 44}.
\bibitem[{of~Medicine(1995)}]{1995-Med}
\bibinfo{author}{I.~of~Medicine}, \bibinfo{title}{Isotopes for Medicine and the
  Life Sciences}, \bibinfo{publisher}{The National Academies Press},
  \bibinfo{address}{Washington, DC}, \bibinfo{year}{1995}.
\bibitem[{Zhuikov(2014)}]{2014-Zhuikov}
\bibinfo{author}{B.~Zhuikov},
\newblock \bibinfo{title}{Production of medical radionuclides in russia: Status
  and future—a review},
\newblock \bibinfo{journal}{Applied Radiation and Isotopes}
  \bibinfo{volume}{84} (\bibinfo{year}{2014}) \bibinfo{pages}{48 -- 56}.
\bibitem[{Hamm and Hamm(2012)}]{2012-Hamm}
\bibinfo{author}{R.~W. Hamm}, \bibinfo{author}{M.~E. Hamm},
  \bibinfo{title}{Industrial Accelerators and Their Applications},
  \bibinfo{publisher}{WORLD SCIENTIFIC}, \bibinfo{year}{2012}.
\bibitem[{Abel et~al.(2019)Abel, Avilov, Ayres, Birnbaum, Bollen, Bonito,
  Bredeweg, Clause, Couture, DeVore, Dietrich, Ellison, Engle, Ferrieri,
  Fitzsimmons, Friedman, Georgobiani, Graves, Greene, Lapi, Loveless, Mastren,
  Martinez-Gomez, McGuinness, Mittig, Morrissey, Peaslee, Pellemoine,
  Robertson, Scielzo, Scott, Severin, Shaughnessy, Shusterman, Singh, Stoyer,
  Sutherlin, Visser, and Wilkinson}]{Abel19}
\bibinfo{author}{E.~P. Abel}, \bibinfo{author}{M.~Avilov},
  \bibinfo{author}{V.~Ayres}, \bibinfo{author}{E.~Birnbaum},
  \bibinfo{author}{G.~Bollen}, \bibinfo{author}{G.~Bonito},
  \bibinfo{author}{T.~Bredeweg}, \bibinfo{author}{H.~Clause},
  \bibinfo{author}{A.~Couture}, \bibinfo{author}{J.~DeVore},
  \bibinfo{author}{M.~Dietrich}, \bibinfo{author}{P.~Ellison},
  \bibinfo{author}{J.~Engle}, \bibinfo{author}{R.~Ferrieri},
  \bibinfo{author}{J.~Fitzsimmons}, \bibinfo{author}{M.~Friedman},
  \bibinfo{author}{D.~Georgobiani}, \bibinfo{author}{S.~Graves},
  \bibinfo{author}{J.~Greene}, \bibinfo{author}{S.~Lapi},
  \bibinfo{author}{C.~S. Loveless}, \bibinfo{author}{T.~Mastren},
  \bibinfo{author}{C.~Martinez-Gomez}, \bibinfo{author}{S.~McGuinness},
  \bibinfo{author}{W.~Mittig}, \bibinfo{author}{D.~Morrissey},
  \bibinfo{author}{G.~Peaslee}, \bibinfo{author}{F.~Pellemoine},
  \bibinfo{author}{J.~D. Robertson}, \bibinfo{author}{N.~Scielzo},
  \bibinfo{author}{M.~Scott}, \bibinfo{author}{G.~Severin},
  \bibinfo{author}{D.~Shaughnessy}, \bibinfo{author}{J.~Shusterman},
  \bibinfo{author}{J.~Singh}, \bibinfo{author}{M.~Stoyer},
  \bibinfo{author}{L.~Sutherlin}, \bibinfo{author}{A.~Visser},
  \bibinfo{author}{J.~Wilkinson},
\newblock \bibinfo{title}{Isotope harvesting at {FRIB}: additional
  opportunities for scientific discovery},
\newblock \bibinfo{journal}{Journal of Physics G: Nuclear and Particle Physics}
  \bibinfo{volume}{46} (\bibinfo{year}{2019}) \bibinfo{pages}{100501}.
\bibitem[{González-Alonso et~al.(2019)González-Alonso, Naviliat-Cuncic, and
  Severijns}]{2018-Gonzalez}
\bibinfo{author}{M.~González-Alonso}, \bibinfo{author}{O.~Naviliat-Cuncic},
  \bibinfo{author}{N.~Severijns},
\newblock \bibinfo{title}{New physics searches in nuclear and neutron β
  decay},
\newblock \bibinfo{journal}{Progress in Particle and Nuclear Physics}
  \bibinfo{volume}{104} (\bibinfo{year}{2019}) \bibinfo{pages}{165 -- 223}.
\bibitem[{Shidling et~al.(2019)Shidling, Kolhinen, Schroeder, Morgan, Ozmetin,
  and Melconian}]{2019-Shidling}
\bibinfo{author}{P.~Shidling}, \bibinfo{author}{V.~Kolhinen},
  \bibinfo{author}{B.~Schroeder}, \bibinfo{author}{N.~Morgan},
  \bibinfo{author}{A.~Ozmetin}, \bibinfo{author}{D.~Melconian},
\newblock \bibinfo{title}{Tamutrap facility: Penning trap facility for weak
  interaction studies},
\newblock \bibinfo{journal}{Hyperfine Interactions} \bibinfo{volume}{240}
  (\bibinfo{year}{2019}).
\bibitem[{O’Malley et~al.(2020)O’Malley, Brodeur, Burdette, Klimes,
  Valverde, Clark, Savard, Ringle, and Varentsov}]{2020-Omalley}
\bibinfo{author}{P.~O’Malley}, \bibinfo{author}{M.~Brodeur},
  \bibinfo{author}{D.~Burdette}, \bibinfo{author}{J.~Klimes},
  \bibinfo{author}{A.~Valverde}, \bibinfo{author}{J.~Clark},
  \bibinfo{author}{G.~Savard}, \bibinfo{author}{R.~Ringle},
  \bibinfo{author}{V.~Varentsov},
\newblock \bibinfo{title}{Testing the weak interaction using st. benedict at
  the university of notre dame},
\newblock \bibinfo{journal}{Nuclear Instruments and Methods in Physics Research
  Section B: Beam Interactions with Materials and Atoms} \bibinfo{volume}{463}
  (\bibinfo{year}{2020}) \bibinfo{pages}{488 -- 490}.
\bibitem[{Ohayon et~al.(2020)Ohayon, Rahangdale, Parnes, Perelman, Heber, and
  Ron}]{ohayon2020}
\bibinfo{author}{B.~Ohayon}, \bibinfo{author}{H.~Rahangdale},
  \bibinfo{author}{E.~Parnes}, \bibinfo{author}{G.~Perelman},
  \bibinfo{author}{O.~Heber}, \bibinfo{author}{G.~Ron},
\newblock \bibinfo{title}{Decay microscope for trapped neon isotopes},
\newblock \bibinfo{journal}{Phys. Rev. C} \bibinfo{volume}{101}
  (\bibinfo{year}{2020}) \bibinfo{pages}{035501}.
\bibitem[{Ohayon et~al.(2018)Ohayon, Chocron, Hirsh, Glick-Magid, Mishnayot,
  Mukul, Rahangdale, Vaintraub, Heber, Gazit, and Ron}]{Ohayon18}
\bibinfo{author}{B.~Ohayon}, \bibinfo{author}{J.~Chocron},
  \bibinfo{author}{T.~Hirsh}, \bibinfo{author}{A.~Glick-Magid},
  \bibinfo{author}{Y.~Mishnayot}, \bibinfo{author}{I.~Mukul},
  \bibinfo{author}{H.~Rahangdale}, \bibinfo{author}{S.~Vaintraub},
  \bibinfo{author}{O.~Heber}, \bibinfo{author}{D.~Gazit},
  \bibinfo{author}{G.~Ron},
\newblock \bibinfo{title}{Weak interaction studies at saraf},
\newblock \bibinfo{journal}{Hyperfine Interactions} \bibinfo{volume}{239}
  (\bibinfo{year}{2018}).
\bibitem[{{Mardor, Israel} et~al.(2018){Mardor, Israel}, {Aviv, Ofer},
  {Avrigeanu, Marilena}, {Berkovits, Dan}, {Dahan, Adi}, {Dickel, Timo},
  {Eliyahu, Ilan}, {Gai, Moshe}, {Gavish-Segev, Inbal}, {Halfon, Shlomi},
  {Hass, Michael}, {Hirsh, Tsviki}, {Kaiser, Boaz}, {Kijel, Daniel}, {Kreisel,
  Arik}, {Mishnayot, Yonatan}, {Mukul, Ish}, {Ohayon, Ben}, {Paul, Michael},
  {Perry, Amichay}, {Rahangdale, Hitesh}, {Rodnizki, Jacob}, {Ron, Guy},
  {Sasson-Zukran, Revital}, {Shor, Asher}, {Silverman, Ido}, {Tessler, Moshe},
  {Vaintraub, Sergey}, and {Weissman, Leo}}]{Mardor18}
\bibinfo{author}{{Mardor, Israel}}, \bibinfo{author}{{Aviv, Ofer}},
  \bibinfo{author}{{Avrigeanu, Marilena}}, \bibinfo{author}{{Berkovits, Dan}},
  \bibinfo{author}{{Dahan, Adi}}, \bibinfo{author}{{Dickel, Timo}},
  \bibinfo{author}{{Eliyahu, Ilan}}, \bibinfo{author}{{Gai, Moshe}},
  \bibinfo{author}{{Gavish-Segev, Inbal}}, \bibinfo{author}{{Halfon, Shlomi}},
  \bibinfo{author}{{Hass, Michael}}, \bibinfo{author}{{Hirsh, Tsviki}},
  \bibinfo{author}{{Kaiser, Boaz}}, \bibinfo{author}{{Kijel, Daniel}},
  \bibinfo{author}{{Kreisel, Arik}}, \bibinfo{author}{{Mishnayot, Yonatan}},
  \bibinfo{author}{{Mukul, Ish}}, \bibinfo{author}{{Ohayon, Ben}},
  \bibinfo{author}{{Paul, Michael}}, \bibinfo{author}{{Perry, Amichay}},
  \bibinfo{author}{{Rahangdale, Hitesh}}, \bibinfo{author}{{Rodnizki, Jacob}},
  \bibinfo{author}{{Ron, Guy}}, \bibinfo{author}{{Sasson-Zukran, Revital}},
  \bibinfo{author}{{Shor, Asher}}, \bibinfo{author}{{Silverman, Ido}},
  \bibinfo{author}{{Tessler, Moshe}}, \bibinfo{author}{{Vaintraub, Sergey}},
  \bibinfo{author}{{Weissman, Leo}},
\newblock \bibinfo{title}{The soreq applied research accelerator facility
  (saraf): Overview, research programs and future plans},
\newblock \bibinfo{journal}{Eur. Phys. J. A} \bibinfo{volume}{54}
  (\bibinfo{year}{2018}) \bibinfo{pages}{91}.
\bibitem[{Hass et~al.(2011)Hass, Vaintraub, Aviv, Blaum, Heber, Mardor,
  Rappaport, Wolf, and Zajfman}]{2011-EIBT}
\bibinfo{author}{M.~Hass}, \bibinfo{author}{S.~Vaintraub},
  \bibinfo{author}{O.~Aviv}, \bibinfo{author}{K.~Blaum},
  \bibinfo{author}{O.~Heber}, \bibinfo{author}{I.~Mardor},
  \bibinfo{author}{M.~Rappaport}, \bibinfo{author}{A.~Wolf},
  \bibinfo{author}{D.~Zajfman},
\newblock \bibinfo{title}{A novel method for fundamental interaction studies
  with an electrostatic ion beam trap},
\newblock in: \bibinfo{booktitle}{Journal of Physics: Conference Series},
  volume \bibinfo{volume}{267}, \bibinfo{organization}{IOP Publishing}, p.
  \bibinfo{pages}{012013}.
\bibitem[{Mukul et~al.(2018)Mukul, Hass, Heber, Hirsh, Mishnayot, Rappaport,
  Ron, Shachar, and Vaintraub}]{Mukul18}
\bibinfo{author}{I.~Mukul}, \bibinfo{author}{M.~Hass},
  \bibinfo{author}{O.~Heber}, \bibinfo{author}{T.~Hirsh},
  \bibinfo{author}{Y.~Mishnayot}, \bibinfo{author}{M.~Rappaport},
  \bibinfo{author}{G.~Ron}, \bibinfo{author}{Y.~Shachar},
  \bibinfo{author}{S.~Vaintraub},
\newblock \bibinfo{title}{A 6he production facility and an electrostatic trap
  for measurement of the beta–neutrino correlation},
\newblock \bibinfo{journal}{Nuclear Instruments and Methods in Physics Research
  Section A: Accelerators, Spectrometers, Detectors and Associated Equipment}
  \bibinfo{volume}{899} (\bibinfo{year}{2018}) \bibinfo{pages}{16 -- 21}.
\bibitem[{Ohayon and Ron(2015)}]{2015-ZS}
\bibinfo{author}{B.~Ohayon}, \bibinfo{author}{G.~Ron},
\newblock \bibinfo{title}{Investigation of different magnetic field
  configurations using an electrical, modular zeeman slower},
\newblock \bibinfo{journal}{Review of Scientific Instruments}
  \bibinfo{volume}{86} (\bibinfo{year}{2015}) \bibinfo{pages}{103110}.
\bibitem[{Ohayon et~al.(2019)Ohayon, Rahangdale, Chocron, Mishnayot, Kosloff,
  Heber, and Ron}]{2019-Imaging}
\bibinfo{author}{B.~Ohayon}, \bibinfo{author}{H.~Rahangdale},
  \bibinfo{author}{J.~Chocron}, \bibinfo{author}{Y.~Mishnayot},
  \bibinfo{author}{R.~Kosloff}, \bibinfo{author}{O.~Heber},
  \bibinfo{author}{G.~Ron},
\newblock \bibinfo{title}{Imaging recoil ions from optical collisions between
  ultracold, metastable neon isotopes},
\newblock \bibinfo{journal}{Physical review letters} \bibinfo{volume}{123}
  (\bibinfo{year}{2019}) \bibinfo{pages}{063401}.
\bibitem[{Zinner et~al.(2003)Zinner, Spoden, Kraemer, Birkl, and
  Ertmer}]{2003-Birkl}
\bibinfo{author}{M.~Zinner}, \bibinfo{author}{P.~Spoden},
  \bibinfo{author}{T.~Kraemer}, \bibinfo{author}{G.~Birkl},
  \bibinfo{author}{W.~Ertmer},
\newblock \bibinfo{title}{Precision measurement of the metastable
  ${}^{3}{P}_{2}$ lifetime of neon},
\newblock \bibinfo{journal}{Phys. Rev. A} \bibinfo{volume}{67}
  (\bibinfo{year}{2003}) \bibinfo{pages}{010501}.
\bibitem[{Ohayon et~al.(2015)Ohayon, W{\aa}hlin, and Ron}]{2015-RF}
\bibinfo{author}{B.~Ohayon}, \bibinfo{author}{E.~W{\aa}hlin},
  \bibinfo{author}{G.~Ron},
\newblock \bibinfo{title}{Characterization of a metastable neon beam extracted
  from a commercial rf ion source},
\newblock \bibinfo{journal}{Journal of Instrumentation} \bibinfo{volume}{10}
  (\bibinfo{year}{2015}) \bibinfo{pages}{P03009}.
\bibitem[{Gonz\'alez-Alonso and Naviliat-Cuncic(2016)}]{2016-OscarBeta}
\bibinfo{author}{M.~Gonz\'alez-Alonso}, \bibinfo{author}{O.~Naviliat-Cuncic},
\newblock \bibinfo{title}{Kinematic sensitivity to the fierz term of
  $\ensuremath{\beta}$-decay differential spectra},
\newblock \bibinfo{journal}{Phys. Rev. C} \bibinfo{volume}{94}
  (\bibinfo{year}{2016}) \bibinfo{pages}{035503}.
\bibitem[{Feinberg et~al.(2009)Feinberg, Paul, Arenshtam, Berkovits, Kijel,
  Nagler, and Silverman}]{2009-LiLIT}
\bibinfo{author}{G.~Feinberg}, \bibinfo{author}{M.~Paul},
  \bibinfo{author}{A.~Arenshtam}, \bibinfo{author}{D.~Berkovits},
  \bibinfo{author}{D.~Kijel}, \bibinfo{author}{A.~Nagler},
  \bibinfo{author}{I.~Silverman},
\newblock \bibinfo{title}{Lilit-a liquid-lithium target as an intense neutron
  source for nuclear astrophysics at the soreq applied research accelerator
  facility},
\newblock \bibinfo{journal}{Nuclear Physics A} \bibinfo{volume}{827}
  (\bibinfo{year}{2009}) \bibinfo{pages}{590c--592c}.
\bibitem[{Silverman et~al.(2018)Silverman, Arenshtam, Berkovits, Eliyahu,
  Gavish, Grin, Halfon, Hass, Hirsh, Kaizer et~al.}]{2018-LILIT}
\bibinfo{author}{I.~Silverman}, \bibinfo{author}{A.~Arenshtam},
  \bibinfo{author}{D.~Berkovits}, \bibinfo{author}{I.~Eliyahu},
  \bibinfo{author}{I.~Gavish}, \bibinfo{author}{A.~Grin},
  \bibinfo{author}{S.~Halfon}, \bibinfo{author}{M.~Hass},
  \bibinfo{author}{T.~Hirsh}, \bibinfo{author}{B.~Kaizer}, et~al.,
\newblock \bibinfo{title}{Scientific opportunities at saraf with a liquid
  lithium jet target neutron source},
\newblock in: \bibinfo{booktitle}{AIP conference proceedings}, volume
  \bibinfo{volume}{1962}, \bibinfo{organization}{AIP Publishing}, p.
  \bibinfo{pages}{020002}.
\bibitem[{Knecht et~al.(2012)}]{2012-Knecht}
\bibinfo{author}{A.~Knecht}, et~al.,
\newblock \bibinfo{title}{{Precision measurement of the He-6 half-life and the
  weak axial current in nuclei}},
\newblock \bibinfo{journal}{Phys. Rev. C} \bibinfo{volume}{86}
  (\bibinfo{year}{2012}) \bibinfo{pages}{035506}.
\bibitem[{Laffoley et~al.(2015)Laffoley, Svensson, Andreoiu, Ball, Bender,
  Bidaman, Bildstein, Blank, Cross, Deng, Varela, Dunlop, Dunlop, Garnsworthy,
  Garrett, Giovinazzo, Grinyer, Grinyer, Hackman, Hadinia, Jamieson,
  Jigmeddorj, Kisliuk, Leach, Leslie, MacLean, Miller, Mills, Moukaddam,
  Radich, Rajabali, Rand, Thomas, Turko, Unsworth, and Voss}]{2015-Laffoley}
\bibinfo{author}{A.~T. Laffoley}, \bibinfo{author}{C.~E. Svensson},
  \bibinfo{author}{C.~Andreoiu}, \bibinfo{author}{G.~C. Ball},
  \bibinfo{author}{P.~C. Bender}, \bibinfo{author}{H.~Bidaman},
  \bibinfo{author}{V.~Bildstein}, \bibinfo{author}{B.~Blank},
  \bibinfo{author}{D.~S. Cross}, \bibinfo{author}{G.~Deng},
  \bibinfo{author}{A.~D. Varela}, \bibinfo{author}{M.~R. Dunlop},
  \bibinfo{author}{R.~Dunlop}, \bibinfo{author}{A.~B. Garnsworthy},
  \bibinfo{author}{P.~E. Garrett}, \bibinfo{author}{J.~Giovinazzo},
  \bibinfo{author}{G.~F. Grinyer}, \bibinfo{author}{J.~Grinyer},
  \bibinfo{author}{G.~Hackman}, \bibinfo{author}{B.~Hadinia},
  \bibinfo{author}{D.~S. Jamieson}, \bibinfo{author}{B.~Jigmeddorj},
  \bibinfo{author}{D.~Kisliuk}, \bibinfo{author}{K.~G. Leach},
  \bibinfo{author}{J.~R. Leslie}, \bibinfo{author}{A.~D. MacLean},
  \bibinfo{author}{D.~Miller}, \bibinfo{author}{B.~Mills},
  \bibinfo{author}{M.~Moukaddam}, \bibinfo{author}{A.~J. Radich},
  \bibinfo{author}{M.~M. Rajabali}, \bibinfo{author}{E.~T. Rand},
  \bibinfo{author}{J.~C. Thomas}, \bibinfo{author}{J.~Turko},
  \bibinfo{author}{C.~Unsworth}, \bibinfo{author}{P.~Voss},
\newblock \bibinfo{title}{High-precision half-life measurements for the
  superallowed fermi ${\ensuremath{\beta}}^{+}$ emitter $^{18}\mathrm{Ne}$},
\newblock \bibinfo{journal}{Phys. Rev. C} \bibinfo{volume}{92}
  (\bibinfo{year}{2015}) \bibinfo{pages}{025502}.
\bibitem[{Segal(2014)}]{Tom}
\bibinfo{author}{T.~Segal}, \bibinfo{title}{The Production of $^{23}$Ne via the
  $^{23}$Na (n,p) Reaction}, Master's thesis, The Hebrew University of
  Jerusalem, \bibinfo{address}{Givat Ram}, \bibinfo{year}{2014}.
\bibitem[{Pollard and Watson(1940)}]{1940-Watson}
\bibinfo{author}{E.~Pollard}, \bibinfo{author}{W.~W. Watson},
\newblock \bibinfo{title}{Transmutation of the separated isotopes of neon by
  deuterons},
\newblock \bibinfo{journal}{Physical Review} \bibinfo{volume}{58}
  (\bibinfo{year}{1940}) \bibinfo{pages}{12}.
\bibitem[{{Penning} and {Schmidt}(1957)}]{Penning57}
\bibinfo{author}{J.~R. {Penning}}, \bibinfo{author}{F.~H. {Schmidt}},
\newblock \bibinfo{title}{{Radioactive Decay of Ne$^{23}$}},
\newblock \bibinfo{journal}{Phys. Rev.} \bibinfo{volume}{105}
  (\bibinfo{year}{1957}) \bibinfo{pages}{647--651}.
\bibitem[{Moreh et~al.(1968)Moreh, Balderman, and Gozez}]{Moreh68}
\bibinfo{author}{R.~Moreh}, \bibinfo{author}{J.~Balderman},
  \bibinfo{author}{Y.~Gozez},
\newblock \bibinfo{title}{Decay of 23ne},
\newblock \bibinfo{journal}{Nuclear Physics A} \bibinfo{volume}{107}
  (\bibinfo{year}{1968}) \bibinfo{pages}{236 -- 240}.
\bibitem[{Lancman et~al.(1965)Lancman, Jasiński, Kownacki, and
  Ludziejewski}]{Lancman65}
\bibinfo{author}{H.~Lancman}, \bibinfo{author}{A.~Jasiński},
  \bibinfo{author}{J.~Kownacki}, \bibinfo{author}{J.~Ludziejewski},
\newblock \bibinfo{title}{The decay scheme of ne23},
\newblock \bibinfo{journal}{Nuclear Physics} \bibinfo{volume}{69}
  (\bibinfo{year}{1965}) \bibinfo{pages}{384 -- 400}.
\bibitem[{Bjerge(1937)}]{1937-Bjerge}
\bibinfo{author}{T.~Bjerge},
\newblock \bibinfo{title}{Radioactive neon},
\newblock \bibinfo{journal}{Nature} \bibinfo{volume}{139}
  (\bibinfo{year}{1937}) \bibinfo{pages}{757}.
\bibitem[{Burman et~al.(1959)Burman, Herrmannsfeldt, Allen, and
  Braid}]{1958-Allen}
\bibinfo{author}{R.~L. Burman}, \bibinfo{author}{W.~B. Herrmannsfeldt},
  \bibinfo{author}{J.~S. Allen}, \bibinfo{author}{T.~H. Braid},
\newblock \bibinfo{title}{Electron-neutrino angular correlation in the beta
  decay of neon-23},
\newblock \bibinfo{journal}{Phys. Rev. Lett.} \bibinfo{volume}{2}
  (\bibinfo{year}{1959}) \bibinfo{pages}{9--11}.
\bibitem[{Ridley(1958)}]{1958-Ridely}
\bibinfo{author}{B.~Ridley},
\newblock \bibinfo{title}{The electron-neutrino angular correlation in decay of
  ne23},
\newblock \bibinfo{journal}{Nuclear Physics} \bibinfo{volume}{6}
  (\bibinfo{year}{1958}) \bibinfo{pages}{34 -- 49}.
\bibitem[{Williamson(1961)}]{1961-Williamson}
\bibinfo{author}{C.~F. Williamson},
\newblock \bibinfo{title}{Absolute cross sections of the reactions
  ${\mathrm{na}}^{23}(n, p){\mathrm{ne}}^{23}$ and ${\mathrm{na}}^{23}(n,
  \ensuremath{\alpha}){\mathrm{f}}^{20}$},
\newblock \bibinfo{journal}{Phys. Rev.} \bibinfo{volume}{122}
  (\bibinfo{year}{1961}) \bibinfo{pages}{1877--1882}.
\bibitem[{Hirsh(2012)}]{Tsviki}
\bibinfo{author}{T.~Y. Hirsh}, \bibinfo{title}{Production of 8 Li and 6 He
  radioactive beams in high current deuteron accelerators}, Ph.D. thesis,
  Weizmann Institute of Science, \bibinfo{year}{2012}.
\bibitem[{Mardor and Berkovits(2015)}]{2015-SARAF}
\bibinfo{author}{I.~Mardor}, \bibinfo{author}{D.~Berkovits},
\newblock \bibinfo{title}{The soreq applied research accelerator facility
  (saraf)},
\newblock \bibinfo{journal}{Nuclear Physics News} \bibinfo{volume}{25}
  (\bibinfo{year}{2015}) \bibinfo{pages}{16--22}.
\bibitem[{Hirsh et~al.(2012)Hirsh, Berkovits, Hass, Jardin, Pichard, Rappaport,
  Shachar, and Silverman}]{2012-LiFTiT}
\bibinfo{author}{T.~Hirsh}, \bibinfo{author}{D.~Berkovits},
  \bibinfo{author}{M.~Hass}, \bibinfo{author}{P.~Jardin},
  \bibinfo{author}{A.~Pichard}, \bibinfo{author}{M.~Rappaport},
  \bibinfo{author}{Y.~Shachar}, \bibinfo{author}{I.~Silverman},
\newblock \bibinfo{title}{Towards an intense radioactive 8li beam at saraf
  phase i},
\newblock in: \bibinfo{booktitle}{Journal of Physics: Conference Series},
  volume \bibinfo{volume}{337}, \bibinfo{organization}{IOP Publishing}, p.
  \bibinfo{pages}{012010}.
\bibitem[{Weissman et~al.(2009)Weissman, Berkovits, Eisen, Halfon, Mardor,
  Perry, and Rodnizki}]{2009-Weissman}
\bibinfo{author}{L.~Weissman}, \bibinfo{author}{D.~Berkovits},
  \bibinfo{author}{Y.~Eisen}, \bibinfo{author}{S.~Halfon},
  \bibinfo{author}{I.~Mardor}, \bibinfo{author}{A.~Perry},
  \bibinfo{author}{J.~Rodnizki},
\newblock \bibinfo{title}{First experience at saraf with proton beams using the
  rutherford scattering monitor},
\newblock \bibinfo{journal}{DIPAC 2009 - 9th European Workshop on Beam
  Diagnostics and Instrumentation for Particle Accelerators}
  (\bibinfo{year}{2009}).
\bibitem[{Sosa et~al.(2017)Sosa, Huber, Welk, and Fraser}]{2017_Sosa}
\bibinfo{author}{J.~Sosa}, \bibinfo{author}{D.~Huber},
  \bibinfo{author}{B.~Welk}, \bibinfo{author}{H.~Fraser},
\newblock \bibinfo{title}{Mipar\textsuperscript{TM}: 2d and 3d image analysis
  software designed by materials scientists, for all scientists},
\newblock \bibinfo{journal}{Microscopy and Microanalysis} \bibinfo{volume}{23}
  (\bibinfo{year}{2017}) \bibinfo{pages}{230--231}.
\bibitem[{Agostinelli et~al.(2003)Agostinelli, Allison, Amako, Apostolakis,
  Araujo, Arce, Asai, Axen, Banerjee, Barrand, Behner, Bellagamba, Boudreau,
  Broglia, Brunengo, Burkhardt, Chauvie, Chuma, Chytracek, Cooperman, Cosmo,
  Degtyarenko, Dell'Acqua, Depaola, Dietrich, Enami, Feliciello, Ferguson,
  Fesefeldt, Folger, Foppiano, Forti, Garelli, Giani, Giannitrapani, Gibin,
  Cadenas], González, Abril], Greeniaus, Greiner, Grichine, Grossheim,
  Guatelli, Gumplinger, Hamatsu, Hashimoto, Hasui, Heikkinen, Howard,
  Ivanchenko, Johnson, Jones, Kallenbach, Kanaya, Kawabata, Kawabata, Kawaguti,
  Kelner, Kent, Kimura, Kodama, Kokoulin, Kossov, Kurashige, Lamanna, Lampén,
  Lara, Lefebure, Lei, Liendl, Lockman, Longo, Magni, Maire, Medernach,
  Minamimoto, de~Freitas], Morita, Murakami, Nagamatu, Nartallo, Nieminen,
  Nishimura, Ohtsubo, Okamura, O'Neale, Oohata, Paech, Perl, Pfeiffer, Pia,
  Ranjard, Rybin, Sadilov, Salvo], Santin, Sasaki, Savvas, Sawada, Scherer,
  Sei, Sirotenko, Smith, Starkov, Stoecker, Sulkimo, Takahata, Tanaka,
  Tcherniaev, Tehrani], Tropeano, Truscott, Uno, Urban, Urban, Verderi,
  Walkden, Wander, Weber, Wellisch, Wenaus, Williams, Wright, Yamada, Yoshida,
  and Zschiesche}]{Geant4a}
\bibinfo{author}{S.~Agostinelli}, \bibinfo{author}{J.~Allison},
  \bibinfo{author}{K.~Amako}, \bibinfo{author}{J.~Apostolakis},
  \bibinfo{author}{H.~Araujo}, \bibinfo{author}{P.~Arce},
  \bibinfo{author}{M.~Asai}, \bibinfo{author}{D.~Axen},
  \bibinfo{author}{S.~Banerjee}, \bibinfo{author}{G.~Barrand},
  \bibinfo{author}{F.~Behner}, \bibinfo{author}{L.~Bellagamba},
  \bibinfo{author}{J.~Boudreau}, \bibinfo{author}{L.~Broglia},
  \bibinfo{author}{A.~Brunengo}, \bibinfo{author}{H.~Burkhardt},
  \bibinfo{author}{S.~Chauvie}, \bibinfo{author}{J.~Chuma},
  \bibinfo{author}{R.~Chytracek}, \bibinfo{author}{G.~Cooperman},
  \bibinfo{author}{G.~Cosmo}, \bibinfo{author}{P.~Degtyarenko},
  \bibinfo{author}{A.~Dell'Acqua}, \bibinfo{author}{G.~Depaola},
  \bibinfo{author}{D.~Dietrich}, \bibinfo{author}{R.~Enami},
  \bibinfo{author}{A.~Feliciello}, \bibinfo{author}{C.~Ferguson},
  \bibinfo{author}{H.~Fesefeldt}, \bibinfo{author}{G.~Folger},
  \bibinfo{author}{F.~Foppiano}, \bibinfo{author}{A.~Forti},
  \bibinfo{author}{S.~Garelli}, \bibinfo{author}{S.~Giani},
  \bibinfo{author}{R.~Giannitrapani}, \bibinfo{author}{D.~Gibin},
  \bibinfo{author}{J.~G. Cadenas]}, \bibinfo{author}{I.~González},
  \bibinfo{author}{G.~G. Abril]}, \bibinfo{author}{G.~Greeniaus},
  \bibinfo{author}{W.~Greiner}, \bibinfo{author}{V.~Grichine},
  \bibinfo{author}{A.~Grossheim}, \bibinfo{author}{S.~Guatelli},
  \bibinfo{author}{P.~Gumplinger}, \bibinfo{author}{R.~Hamatsu},
  \bibinfo{author}{K.~Hashimoto}, \bibinfo{author}{H.~Hasui},
  \bibinfo{author}{A.~Heikkinen}, \bibinfo{author}{A.~Howard},
  \bibinfo{author}{V.~Ivanchenko}, \bibinfo{author}{A.~Johnson},
  \bibinfo{author}{F.~Jones}, \bibinfo{author}{J.~Kallenbach},
  \bibinfo{author}{N.~Kanaya}, \bibinfo{author}{M.~Kawabata},
  \bibinfo{author}{Y.~Kawabata}, \bibinfo{author}{M.~Kawaguti},
  \bibinfo{author}{S.~Kelner}, \bibinfo{author}{P.~Kent},
  \bibinfo{author}{A.~Kimura}, \bibinfo{author}{T.~Kodama},
  \bibinfo{author}{R.~Kokoulin}, \bibinfo{author}{M.~Kossov},
  \bibinfo{author}{H.~Kurashige}, \bibinfo{author}{E.~Lamanna},
  \bibinfo{author}{T.~Lampén}, \bibinfo{author}{V.~Lara},
  \bibinfo{author}{V.~Lefebure}, \bibinfo{author}{F.~Lei},
  \bibinfo{author}{M.~Liendl}, \bibinfo{author}{W.~Lockman},
  \bibinfo{author}{F.~Longo}, \bibinfo{author}{S.~Magni},
  \bibinfo{author}{M.~Maire}, \bibinfo{author}{E.~Medernach},
  \bibinfo{author}{K.~Minamimoto}, \bibinfo{author}{P.~M. de~Freitas]},
  \bibinfo{author}{Y.~Morita}, \bibinfo{author}{K.~Murakami},
  \bibinfo{author}{M.~Nagamatu}, \bibinfo{author}{R.~Nartallo},
  \bibinfo{author}{P.~Nieminen}, \bibinfo{author}{T.~Nishimura},
  \bibinfo{author}{K.~Ohtsubo}, \bibinfo{author}{M.~Okamura},
  \bibinfo{author}{S.~O'Neale}, \bibinfo{author}{Y.~Oohata},
  \bibinfo{author}{K.~Paech}, \bibinfo{author}{J.~Perl},
  \bibinfo{author}{A.~Pfeiffer}, \bibinfo{author}{M.~Pia},
  \bibinfo{author}{F.~Ranjard}, \bibinfo{author}{A.~Rybin},
  \bibinfo{author}{S.~Sadilov}, \bibinfo{author}{E.~D. Salvo]},
  \bibinfo{author}{G.~Santin}, \bibinfo{author}{T.~Sasaki},
  \bibinfo{author}{N.~Savvas}, \bibinfo{author}{Y.~Sawada},
  \bibinfo{author}{S.~Scherer}, \bibinfo{author}{S.~Sei},
  \bibinfo{author}{V.~Sirotenko}, \bibinfo{author}{D.~Smith},
  \bibinfo{author}{N.~Starkov}, \bibinfo{author}{H.~Stoecker},
  \bibinfo{author}{J.~Sulkimo}, \bibinfo{author}{M.~Takahata},
  \bibinfo{author}{S.~Tanaka}, \bibinfo{author}{E.~Tcherniaev},
  \bibinfo{author}{E.~S. Tehrani]}, \bibinfo{author}{M.~Tropeano},
  \bibinfo{author}{P.~Truscott}, \bibinfo{author}{H.~Uno},
  \bibinfo{author}{L.~Urban}, \bibinfo{author}{P.~Urban},
  \bibinfo{author}{M.~Verderi}, \bibinfo{author}{A.~Walkden},
  \bibinfo{author}{W.~Wander}, \bibinfo{author}{H.~Weber},
  \bibinfo{author}{J.~Wellisch}, \bibinfo{author}{T.~Wenaus},
  \bibinfo{author}{D.~Williams}, \bibinfo{author}{D.~Wright},
  \bibinfo{author}{T.~Yamada}, \bibinfo{author}{H.~Yoshida},
  \bibinfo{author}{D.~Zschiesche},
\newblock \bibinfo{title}{Geant4—a simulation toolkit},
\newblock \bibinfo{journal}{Nuclear Instruments and Methods in Physics Research
  Section A: Accelerators, Spectrometers, Detectors and Associated Equipment}
  \bibinfo{volume}{506} (\bibinfo{year}{2003}) \bibinfo{pages}{250 -- 303}.
\bibitem[{{Allison} et~al.(2006){Allison}, {Amako}, {Apostolakis}, {Araujo},
  {Arce Dubois}, {Asai}, {Barrand}, {Capra}, {Chauvie}, {Chytracek}, {Cirrone},
  {Cooperman}, {Cosmo}, {Cuttone}, {Daquino}, {Donszelmann}, {Dressel},
  {Folger}, {Foppiano}, {Generowicz}, {Grichine}, {Guatelli}, {Gumplinger},
  {Heikkinen}, {Hrivnacova}, {Howard}, {Incerti}, {Ivanchenko}, {Johnson},
  {Jones}, {Koi}, {Kokoulin}, {Kossov}, {Kurashige}, {Lara}, {Larsson}, {Lei},
  {Link}, {Longo}, {Maire}, {Mantero}, {Mascialino}, {McLaren}, {Mendez
  Lorenzo}, {Minamimoto}, {Murakami}, {Nieminen}, {Pandola}, {Parlati},
  {Peralta}, {Perl}, {Pfeiffer}, {Pia}, {Ribon}, {Rodrigues}, {Russo},
  {Sadilov}, {Santin}, {Sasaki}, {Smith}, {Starkov}, {Tanaka}, {Tcherniaev},
  {Tome}, {Trindade}, {Truscott}, {Urban}, {Verderi}, {Walkden}, {Wellisch},
  {Williams}, {Wright}, and {Yoshida}}]{Geant4b}
\bibinfo{author}{J.~{Allison}}, \bibinfo{author}{K.~{Amako}},
  \bibinfo{author}{J.~{Apostolakis}}, \bibinfo{author}{H.~{Araujo}},
  \bibinfo{author}{P.~{Arce Dubois}}, \bibinfo{author}{M.~{Asai}},
  \bibinfo{author}{G.~{Barrand}}, \bibinfo{author}{R.~{Capra}},
  \bibinfo{author}{S.~{Chauvie}}, \bibinfo{author}{R.~{Chytracek}},
  \bibinfo{author}{G.~A.~P. {Cirrone}}, \bibinfo{author}{G.~{Cooperman}},
  \bibinfo{author}{G.~{Cosmo}}, \bibinfo{author}{G.~{Cuttone}},
  \bibinfo{author}{G.~G. {Daquino}}, \bibinfo{author}{M.~{Donszelmann}},
  \bibinfo{author}{M.~{Dressel}}, \bibinfo{author}{G.~{Folger}},
  \bibinfo{author}{F.~{Foppiano}}, \bibinfo{author}{J.~{Generowicz}},
  \bibinfo{author}{V.~{Grichine}}, \bibinfo{author}{S.~{Guatelli}},
  \bibinfo{author}{P.~{Gumplinger}}, \bibinfo{author}{A.~{Heikkinen}},
  \bibinfo{author}{I.~{Hrivnacova}}, \bibinfo{author}{A.~{Howard}},
  \bibinfo{author}{S.~{Incerti}}, \bibinfo{author}{V.~{Ivanchenko}},
  \bibinfo{author}{T.~{Johnson}}, \bibinfo{author}{F.~{Jones}},
  \bibinfo{author}{T.~{Koi}}, \bibinfo{author}{R.~{Kokoulin}},
  \bibinfo{author}{M.~{Kossov}}, \bibinfo{author}{H.~{Kurashige}},
  \bibinfo{author}{V.~{Lara}}, \bibinfo{author}{S.~{Larsson}},
  \bibinfo{author}{F.~{Lei}}, \bibinfo{author}{O.~{Link}},
  \bibinfo{author}{F.~{Longo}}, \bibinfo{author}{M.~{Maire}},
  \bibinfo{author}{A.~{Mantero}}, \bibinfo{author}{B.~{Mascialino}},
  \bibinfo{author}{I.~{McLaren}}, \bibinfo{author}{P.~{Mendez Lorenzo}},
  \bibinfo{author}{K.~{Minamimoto}}, \bibinfo{author}{K.~{Murakami}},
  \bibinfo{author}{P.~{Nieminen}}, \bibinfo{author}{L.~{Pandola}},
  \bibinfo{author}{S.~{Parlati}}, \bibinfo{author}{L.~{Peralta}},
  \bibinfo{author}{J.~{Perl}}, \bibinfo{author}{A.~{Pfeiffer}},
  \bibinfo{author}{M.~G. {Pia}}, \bibinfo{author}{A.~{Ribon}},
  \bibinfo{author}{P.~{Rodrigues}}, \bibinfo{author}{G.~{Russo}},
  \bibinfo{author}{S.~{Sadilov}}, \bibinfo{author}{G.~{Santin}},
  \bibinfo{author}{T.~{Sasaki}}, \bibinfo{author}{D.~{Smith}},
  \bibinfo{author}{N.~{Starkov}}, \bibinfo{author}{S.~{Tanaka}},
  \bibinfo{author}{E.~{Tcherniaev}}, \bibinfo{author}{B.~{Tome}},
  \bibinfo{author}{A.~{Trindade}}, \bibinfo{author}{P.~{Truscott}},
  \bibinfo{author}{L.~{Urban}}, \bibinfo{author}{M.~{Verderi}},
  \bibinfo{author}{A.~{Walkden}}, \bibinfo{author}{J.~P. {Wellisch}},
  \bibinfo{author}{D.~C. {Williams}}, \bibinfo{author}{D.~{Wright}},
  \bibinfo{author}{H.~{Yoshida}},
\newblock \bibinfo{title}{Geant4 developments and applications},
\newblock \bibinfo{journal}{IEEE Transactions on Nuclear Science}
  \bibinfo{volume}{53} (\bibinfo{year}{2006}) \bibinfo{pages}{270--278}.
\bibitem[{Allison et~al.(2016)Allison, Amako, Apostolakis, Arce, Asai, Aso,
  Bagli, Bagulya, Banerjee, Barrand, Beck, Bogdanov, Brandt, Brown, Burkhardt,
  Canal, Cano-Ott, Chauvie, Cho, Cirrone, Cooperman, Cortés-Giraldo, Cosmo,
  Cuttone, Depaola, Desorgher, Dong, Dotti, Elvira, Folger, Francis, Galoyan,
  Garnier, Gayer, Genser, Grichine, Guatelli, Guèye, Gumplinger, Howard,
  Hřivnáčová, Hwang, Incerti, Ivanchenko, Ivanchenko, Jones, Jun,
  Kaitaniemi, Karakatsanis, Karamitros, Kelsey, Kimura, Koi, Kurashige,
  Lechner, Lee, Longo, Maire, Mancusi, Mantero, Mendoza, Morgan, Murakami,
  Nikitina, Pandola, Paprocki, Perl, Petrović, Pia, Pokorski, Quesada, Raine,
  Reis, Ribon, Fira], Romano, Russo, Santin, Sasaki, Sawkey, Shin, Strakovsky,
  Taborda, Tanaka, Tomé, Toshito, Tran, Truscott, Urban, Uzhinsky, Verbeke,
  Verderi, Wendt, Wenzel, Wright, Wright, Yamashita, Yarba, and
  Yoshida}]{Geant4c}
\bibinfo{author}{J.~Allison}, \bibinfo{author}{K.~Amako},
  \bibinfo{author}{J.~Apostolakis}, \bibinfo{author}{P.~Arce},
  \bibinfo{author}{M.~Asai}, \bibinfo{author}{T.~Aso},
  \bibinfo{author}{E.~Bagli}, \bibinfo{author}{A.~Bagulya},
  \bibinfo{author}{S.~Banerjee}, \bibinfo{author}{G.~Barrand},
  \bibinfo{author}{B.~Beck}, \bibinfo{author}{A.~Bogdanov},
  \bibinfo{author}{D.~Brandt}, \bibinfo{author}{J.~Brown},
  \bibinfo{author}{H.~Burkhardt}, \bibinfo{author}{P.~Canal},
  \bibinfo{author}{D.~Cano-Ott}, \bibinfo{author}{S.~Chauvie},
  \bibinfo{author}{K.~Cho}, \bibinfo{author}{G.~Cirrone},
  \bibinfo{author}{G.~Cooperman}, \bibinfo{author}{M.~Cortés-Giraldo},
  \bibinfo{author}{G.~Cosmo}, \bibinfo{author}{G.~Cuttone},
  \bibinfo{author}{G.~Depaola}, \bibinfo{author}{L.~Desorgher},
  \bibinfo{author}{X.~Dong}, \bibinfo{author}{A.~Dotti},
  \bibinfo{author}{V.~Elvira}, \bibinfo{author}{G.~Folger},
  \bibinfo{author}{Z.~Francis}, \bibinfo{author}{A.~Galoyan},
  \bibinfo{author}{L.~Garnier}, \bibinfo{author}{M.~Gayer},
  \bibinfo{author}{K.~Genser}, \bibinfo{author}{V.~Grichine},
  \bibinfo{author}{S.~Guatelli}, \bibinfo{author}{P.~Guèye},
  \bibinfo{author}{P.~Gumplinger}, \bibinfo{author}{A.~Howard},
  \bibinfo{author}{I.~Hřivnáčová}, \bibinfo{author}{S.~Hwang},
  \bibinfo{author}{S.~Incerti}, \bibinfo{author}{A.~Ivanchenko},
  \bibinfo{author}{V.~Ivanchenko}, \bibinfo{author}{F.~Jones},
  \bibinfo{author}{S.~Jun}, \bibinfo{author}{P.~Kaitaniemi},
  \bibinfo{author}{N.~Karakatsanis}, \bibinfo{author}{M.~Karamitros},
  \bibinfo{author}{M.~Kelsey}, \bibinfo{author}{A.~Kimura},
  \bibinfo{author}{T.~Koi}, \bibinfo{author}{H.~Kurashige},
  \bibinfo{author}{A.~Lechner}, \bibinfo{author}{S.~Lee},
  \bibinfo{author}{F.~Longo}, \bibinfo{author}{M.~Maire},
  \bibinfo{author}{D.~Mancusi}, \bibinfo{author}{A.~Mantero},
  \bibinfo{author}{E.~Mendoza}, \bibinfo{author}{B.~Morgan},
  \bibinfo{author}{K.~Murakami}, \bibinfo{author}{T.~Nikitina},
  \bibinfo{author}{L.~Pandola}, \bibinfo{author}{P.~Paprocki},
  \bibinfo{author}{J.~Perl}, \bibinfo{author}{I.~Petrović},
  \bibinfo{author}{M.~Pia}, \bibinfo{author}{W.~Pokorski},
  \bibinfo{author}{J.~Quesada}, \bibinfo{author}{M.~Raine},
  \bibinfo{author}{M.~Reis}, \bibinfo{author}{A.~Ribon}, \bibinfo{author}{A.~R.
  Fira]}, \bibinfo{author}{F.~Romano}, \bibinfo{author}{G.~Russo},
  \bibinfo{author}{G.~Santin}, \bibinfo{author}{T.~Sasaki},
  \bibinfo{author}{D.~Sawkey}, \bibinfo{author}{J.~Shin},
  \bibinfo{author}{I.~Strakovsky}, \bibinfo{author}{A.~Taborda},
  \bibinfo{author}{S.~Tanaka}, \bibinfo{author}{B.~Tomé},
  \bibinfo{author}{T.~Toshito}, \bibinfo{author}{H.~Tran},
  \bibinfo{author}{P.~Truscott}, \bibinfo{author}{L.~Urban},
  \bibinfo{author}{V.~Uzhinsky}, \bibinfo{author}{J.~Verbeke},
  \bibinfo{author}{M.~Verderi}, \bibinfo{author}{B.~Wendt},
  \bibinfo{author}{H.~Wenzel}, \bibinfo{author}{D.~Wright},
  \bibinfo{author}{D.~Wright}, \bibinfo{author}{T.~Yamashita},
  \bibinfo{author}{J.~Yarba}, \bibinfo{author}{H.~Yoshida},
\newblock \bibinfo{title}{Recent developments in geant4},
\newblock \bibinfo{journal}{Nuclear Instruments and Methods in Physics Research
  Section A: Accelerators, Spectrometers, Detectors and Associated Equipment}
  \bibinfo{volume}{835} (\bibinfo{year}{2016}) \bibinfo{pages}{186 -- 225}.
\bibitem[{V{\"a}nsk{\"a} and Rieppo(1979)}]{1979-vanska}
\bibinfo{author}{R.~V{\"a}nsk{\"a}}, \bibinfo{author}{R.~Rieppo},
\newblock \bibinfo{title}{Measurement of the average cross sections for (n, p)
  reactions on some light and medium mass nuclei in the 241 ambe neutron
  spectrum},
\newblock \bibinfo{journal}{The International Journal of Applied Radiation and
  Isotopes} \bibinfo{volume}{30} (\bibinfo{year}{1979}) \bibinfo{pages}{513 --
  514}.
\bibitem[{West and Menzies(1929)}]{West29}
\bibinfo{author}{W.~A. West}, \bibinfo{author}{A.~W.~C. Menzies},
\newblock \bibinfo{title}{The vapor pressures of sulphur between 100ֲ° and
  550ֲ° with related thermal data},
\newblock \bibinfo{journal}{The Journal of Physical Chemistry}
  \bibinfo{volume}{33} (\bibinfo{year}{1929}) \bibinfo{pages}{1880--1892}.
\bibitem[{Weigmann et~al.(1983)Weigmann, Auchampaugh, Lisowski, Moore, and
  Morgan}]{1983-Weigmann}
\bibinfo{author}{H.~Weigmann}, \bibinfo{author}{G.~F. Auchampaugh},
  \bibinfo{author}{P.~W. Lisowski}, \bibinfo{author}{M.~S. Moore},
  \bibinfo{author}{G.~L. Morgan},
\newblock \bibinfo{title}{Neutron induced charged particle reactions on 23na},
\newblock in: \bibinfo{editor}{K.~H. B{\"o}ckhoff} (Ed.),
  \bibinfo{booktitle}{Nuclear Data for Science and Technology},
  \bibinfo{publisher}{Springer Netherlands}, \bibinfo{address}{Dordrecht},
  \bibinfo{year}{1983}, pp. \bibinfo{pages}{814--817}.
\bibitem[{{Gerber} et~al.(1956){Gerber}, {Mu{\~n}oz}, and
  {Maeder}}]{1956-Gerber}
\bibinfo{author}{H.~J. {Gerber}}, \bibinfo{author}{M.~G. {Mu{\~n}oz}},
  \bibinfo{author}{D.~{Maeder}},
\newblock \bibinfo{title}{{Beta Decay of Ne$^{23}$}},
\newblock \bibinfo{journal}{Physical Review} \bibinfo{volume}{101}
  (\bibinfo{year}{1956}) \bibinfo{pages}{774--775}.
\bibitem[{Alburger(1974)}]{1974-Alburger}
\bibinfo{author}{D.~E. Alburger},
\newblock \bibinfo{title}{Decays of 21na, 23ne, 23mg, 24ne, and 31s},
\newblock \bibinfo{journal}{Physical Review C} \bibinfo{volume}{9}
  (\bibinfo{year}{1974}) \bibinfo{pages}{991--995}.
\bibitem[{Kersevan and Ady(2019)}]{2019-Kersevan}
\bibinfo{author}{R.~Kersevan}, \bibinfo{author}{M.~Ady},
\newblock \bibinfo{title}{{Recent developments of Monte-Carlo codes Molflow+
  and Synrad+}},
\newblock in: \bibinfo{booktitle}{{10th International Particle Accelerator
  Conference}}, p. \bibinfo{pages}{TUPMP037}.
\bibitem[{Mehrer(2007)}]{Mehrer07}
\bibinfo{author}{H.~Mehrer}, \bibinfo{title}{{Diffusion in Solids:
  Fundamentals, Methods, Materials, Diffusion-Controlled Processes}}, Springer
  Series in Solid-State Sciences, \bibinfo{publisher}{Springer},
  \bibinfo{address}{Berlin, Heidelberg}, \bibinfo{year}{2007}.
\bibitem[{Askeland et~al.(2011)Askeland, Fulay, and Wright}]{Askeland11}
\bibinfo{author}{D.~R. Askeland}, \bibinfo{author}{P.~P. Fulay},
  \bibinfo{author}{W.~J. Wright}, \bibinfo{title}{The Science and Engineering
  of Materials, SI Edition}, \bibinfo{publisher}{CL-Engineering},
  \bibinfo{edition}{6th} edition, \bibinfo{year}{2011}.

\end{thebibliography}
\end{document}